\documentclass{agujournal2019}
\journalname{JGR-Earth Surface}

\usepackage{amsbsy}
\usepackage[sort&compress,square,comma,authoryear]{natbib}

\begin{document}

%
%

\title{Acoustic emissions of nearly steady and uniform granular flows: a proxy for flow dynamics and velocity fluctuations}

%
%

 \authors{Bachelet, V.\affil{1}, Mangeney, A.\affil{1,2}, Toussaint, R.\affil{3,4}, DeRosny, J.\affil{5}, Farin, M.\affil{5}, and Hibert, C.\affil{3}}

\affiliation{1}{Institut de Physique du Globe de Paris, Universit\'{e} Paris-Diderot, Sorbonne Paris Cit\'{e}, CNRS (UMR 7154), 75005 Paris, France}
\affiliation{2}{ANGE team, Inria, Lab. J.-L. Lions, CNRS, 75005, Paris, France}
\affiliation{3}{Institut de Physique du Globe de Strasbourg, Universit\'{e} de Strasbourg/EOST, CNRS, 67000, Strasbourg, France}
\affiliation{4}{PoreLab, Njord Centre, Department of Physics, University of Oslo, Oslo, Norway}
\affiliation{5}{Institut Langevin, ESPCI Paris, CNRS, PSL Research University, 75005, Paris, France}

\correspondingauthor{Toussaint, R.}{renaud.toussaint@unistra.fr}


\begin{keypoints}
\item We analyze the high-frequency emissions and particle agitation of quasi steady granular 
flows on constant slopes.
\item Scaling laws between granular temperature, average velocity, shear rate and inertial number are derived.
\item A simple physical model for the acoustic emissions and acoustic efficiency of steady flows is developed and tested.
\end{keypoints}

%
%

\begin{abstract}
The seismic waves emitted during granular flows are generated by different sources: high frequencies by inter-particle shocks and low frequencies by global motion and large scale deformation. To unravel these different mechanisms, an experimental study has been performed on the seismic waves emitted by dry quasi steady granular flows. The emitted seismic waves were recorded using shock accelerometers and the flow dynamics were captured with a fast camera. 
The mechanical characteristics of the particle shocks were analyzed, along with the duration between shocks and the correlations in the particle motion. The high-frequency seismic waves (1-50 kHz) were found to originate from particle shocks and waves trapped in the flowing layer. The low-frequency waves (20-60 Hz) were generated by the oscillations of the particles along their trajectories, i.e. from cycles of dilation/compression during the shear.
The profiles of granular temperature (i.e. the square of particle velocity fluctuations) and average velocity were measured and related to the average properties of the flow as well as to the slope angle and flow thickness. These profiles were then used in a simple steady granular flow model to predict the radiated seismic energy and the energetic efficiency, i.e. the fraction of the flow potential energy converted to seismic energy. Scaling laws relating the seismic power, the shear strain rate and the inertial number were derived. In particular, the emitted seismic power is proportional to the granular temperature, which is also related to the mean flow velocity. 
\end{abstract}

%
%

\section{Introduction}

Gravitational flows such as landslides, debris avalanches and rockfalls represent one of the major natural hazards threatening life and property in mountainous, volcanic, seismic and coastal areas, with large events possibly displacing several hundred thousand people. They play a key role in erosion processes on the Earth's surface. Gravitational instabilities are also closely related to volcanic, seismic and climatic activity and thus represent potential precursors or proxies for changes in these activities with time, as shown for example for the Piton de la Fournaise volcano, La R\'{e}union Island \citep{Hibert2014, Hibert2017a, Durand2018} or for the Soufri\`{e}re Hills volcano, Montserrat Island \citep{Calder2005}.

Research involving the dynamic analysis of gravitational mass flows is advancing rapidly. One of its ultimate goals is to produce tools for detection of natural instabilities and for prediction of velocity, dynamic pressure and runout extent of rapid landslides. However, the theoretical description and physical understanding of these processes in a natural environment are still open and extremely challenging problems [see \citet{Delannay2017} for a review]. In particular, the origin of the high mobility of large landslides is still unexplained with different hypotheses proposed in the literature (acoustic fluidization, flash heating, etc.) \citep{Lucas2014}. The lack of field measurements relevant to the dynamics of natural landslides prevents us from fully understanding the processes involved and from predicting landslide dynamics and deposition. Indeed, these events are generally unpredictable, but have a strongly destructive power. Furthermore, data on the deposits are not always available due to subsequent flows, erosion processes and site inaccessibility.

In this context, the analysis of the seismic signal generated by natural instabilities provides a unique way to detect and characterize these events and to discriminate between the physical processes involved. When flowing down the slope, landslides generate seismic waves in a wide frequency range that are recorded by local, regional or global seismic networks, depending on the event size \citep{Okal1990,Allstadt2018}.  As a result, the recorded seismic signal, with frequencies ranging from about 0.006 Hz to 30 Hz, carries key information on the landslide dynamics to distances far from the source. However, inferring information from the seismic signal to characterize the landslide source suffers from uncertainties related to the respective effects of the mean flow dynamics, grain-scale processes, topography, mass involved and wave propagation on the recorded signal. It is commonly speculated that grain impacts on the substrate generate high frequencies ($>1$~Hz), while the mean flow acceleration/deceleration is responsible for lower frequencies.

\begin{figure}
\centering
\includegraphics[width=\linewidth]{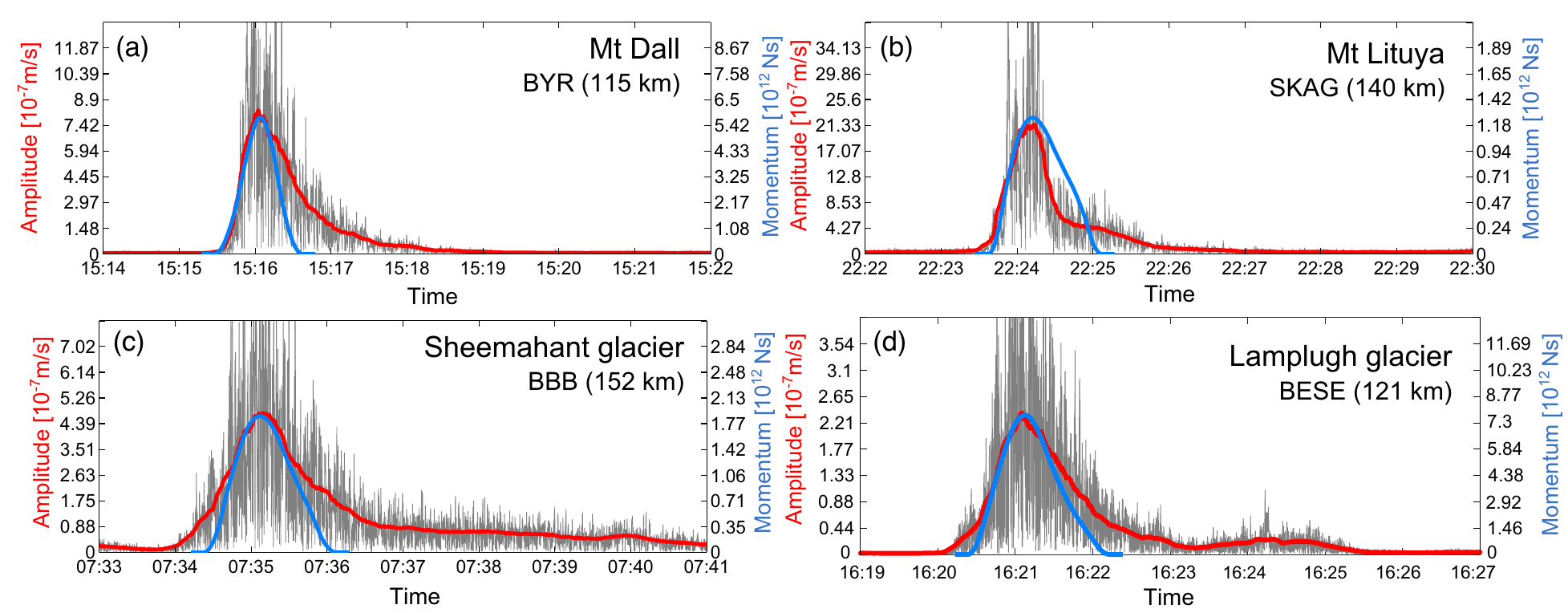}
	\caption{Seismic signal envelope (gray), smoothed envelope (red) and inverted momentum (blue) from the inversion method proposed by \citet{Ekstrom2013} for landslides on a) Mt Dall, b) Mt Lituya, c) the Sheemahant glacier and d) the Lamplugh glacier.}
	\label{fig:terrain}
\end{figure}

Much work has been devoted to extracting information on flow dynamics from the low-frequency signal (periods 10~s$<T<$-120~s) by recovering the force that the landslide applies to the ground from signal deconvolution, e.g. \citet{Kanamori1982, LaRocca2004, Lin2010, Moretti2012, Yamada2013, Allstadt2013,  Ekstrom2013, Zhao2015, Hibert2017b}. The time history of this force is essentially related to the acceleration/deceleration of the flow along the topography. Comparing this force with the force simulated with landslide models makes it possible to recover landslide characteristics and dynamics such as its volume and timing, the friction coefficients involved, the role of erosion processes, bedrock nature (rock or ice) and topography \citep{Favreau2010, Schneider2010, Moretti2012, Moretti2015, Yamada2016, Yamada2018}.

The high-frequency signal is much more difficult to interpret, in particular due to the strong effect of topography and Earth heterogeneity along the path of the seismic waves from the source to the receiver. For these reasons, mainly empirical relationships have been proposed between the high-frequency signal and landslide characteristics \citep{Norris1994, Deparis2008, Dammeier2011}. This high-frequency signal is however more generally recorded because of the lower price of short period seismometers and because small landslides (with volumes $<1000$~m$^3$) only generate frequencies larger than about 1 Hz. Recent studies show correlations between the high-frequency signal (energy, envelope, etc.) and the mean properties of the flow (potential energy lost, force, velocity, momentum, etc.) estimated using landslide models \citep{Hibert2011, Hibert2014, Levy2015} or from inversion of low-frequency seismic data \citep{Hibert2017b}. In particular, \citet{Hibert2017b} observed that the flow momentum is generally proportional to the high-frequency envelope of the signal. However, sometimes, in particular during the deceleration phases, a high-frequency signal can be observed even if the force inverted from the seismic signal, which is proportional to the landslide acceleration, is almost zero, leading to an apparent zero-velocity (see gray area in Fig. \ref{fig:terrain}). However zero acceleration could correspond to constant velocity flows which would generate seismic waves, possibly due to grain agitation. \citet{Huang2007} compared the high-frequency seismic signals generated by rock impacts and debris flows (grain/fluid mixtures) and concluded that one of the main sources of ground vibration caused by debris flows was the interaction of rocks or boulders with the channel bed. However, the complexity of natural landslides and the difficulty to obtain accurate measurements of their dynamics makes it nearly impossible to quantify the link between grain scale physical processes such as velocity fluctuations and the generated seismic signal. More generally, measurements of particle agitation, called granular temperature in the kinetic theory of granular flows, and its link with mean flow properties in dense flows, are still open questions, closely related to the rheology of granular materials [see e.g. \citet{Andreotti2013, Delannay2017} for review papers].

We propose to address this issue here by developing laboratory experiments to record and quantify seismic (i.e. acoustic) waves generated by almost steady and uniform granular flows. Only a few experiments on granular flows and generated acoustic waves have been conducted. These experiments however make it possible to test physical interpretations of the characteristics of the seismic signal generated by natural landslides and to quantify the energy partitioning between the flow and the seismic emissions. Furthermore, such experiments provide a unique way to check models of granular flows and seismic wave generation in a simple configuration, before tackling natural applications. On a 8-meter long channel, \citet{Huang2004} investigated the acoustic waves generated by the friction and  impacts of rocks of about 100 g to 1 kg on a granular bed filled with water and slurry and by a mixture of gravel and water/slurry. They recorded similar frequencies for individual rock motion and debris flows, as observed in the field by \citet{Huang2007}. Their measurements also showed that the amplitude of the acoustic signal increases with the gravel size. However, the complexity of the materials involved and the lack of measurements at the grain scale made it difficult to capture the origin of the generated signal and to quantify the link between the acoustic measurements and the flow properties.

A series of experiments of granular impacts on various beds showed that Hertz theory quantitatively explains the generated acoustic signal on smooth beds \citep{Farin2015}. These experiments also showed that power laws issued from this theory make it possible to relate empirically the acoustic energy to the properties of the impactor (mass, velocity) on smooth, rough and erodible beds \citep{Farin2015, Farin2016, Bachelet2018}. More specifically, the frequency/energy of the acoustic signal is shown to decrease/increase with the mass and velocity of the impactor, respectively, as observed for debris flows \citep{Okuda1980} or for single block rockfalls \citep{Hibert2017c}. Finding these quantitative relationships between acoustic and flow properties was only possible because of the joint accurate measurements and calculations of the grain motion and absolute value of the radiated energy using coupled optical and acoustic methods. In this way, \citet{Farin2018, Farin2019} showed that the seismic power varies in the same manner as the flow velocity in granular collapses on inclined planes. In particular, after the first acceleration/deceleration phase of the mass, the seismic power increases with increasing slope in the same way as the downslope velocity and the agitation of the particle at the flow front. Measurements of grain-scale fluctuations were however not performed in these 3D experiments.

In a quite different setting involving granular materials sheared in a torsional rheometer, \citet{Taylor2017} found that the square of the acceleration measured with their accelerometers divided by the number of particles was proportional to $I \times d^3$, where $d$ is the particle diameter and $I$ the so-called inertial number, defined as the ratio between the time scale related to shear and the time scale related to particle rearrangement under confining pressure. However, \citet{Taylor2017} neither calculated absolute values of the acoustic energy nor measured the characteristics of the flow such as velocity fluctuations, mean velocity profiles, etc.

We investigate here the quantitative link between velocity fluctuations, mean flow properties and acoustic energy by combining accurate optical and acoustic measurements of granular flows over a range of slopes. Our objectives are to: (1) capture and quantify the fluctuations and heterogeneities in almost steady uniform flows and their relationship with mean flow properties, (2) characterize and quantify the radiated acoustic energy, (3) relate the acoustic characteristics (energy, frequency) to the grain scale and mean properties of the flow, (4) check whether a simple model based on particle collisions at fluctuating velocities can explain quantitatively the measured seismic power, (5) quantify the relative contributions of collisions within the flow and with the bed on the generated acoustic energy, (6) quantify the percentage of energy lost by vibrations and (7) discuss our results with regards to field observations.

\section{Set-up}
\begin{figure}
\centering
\includegraphics[width = 10cm]{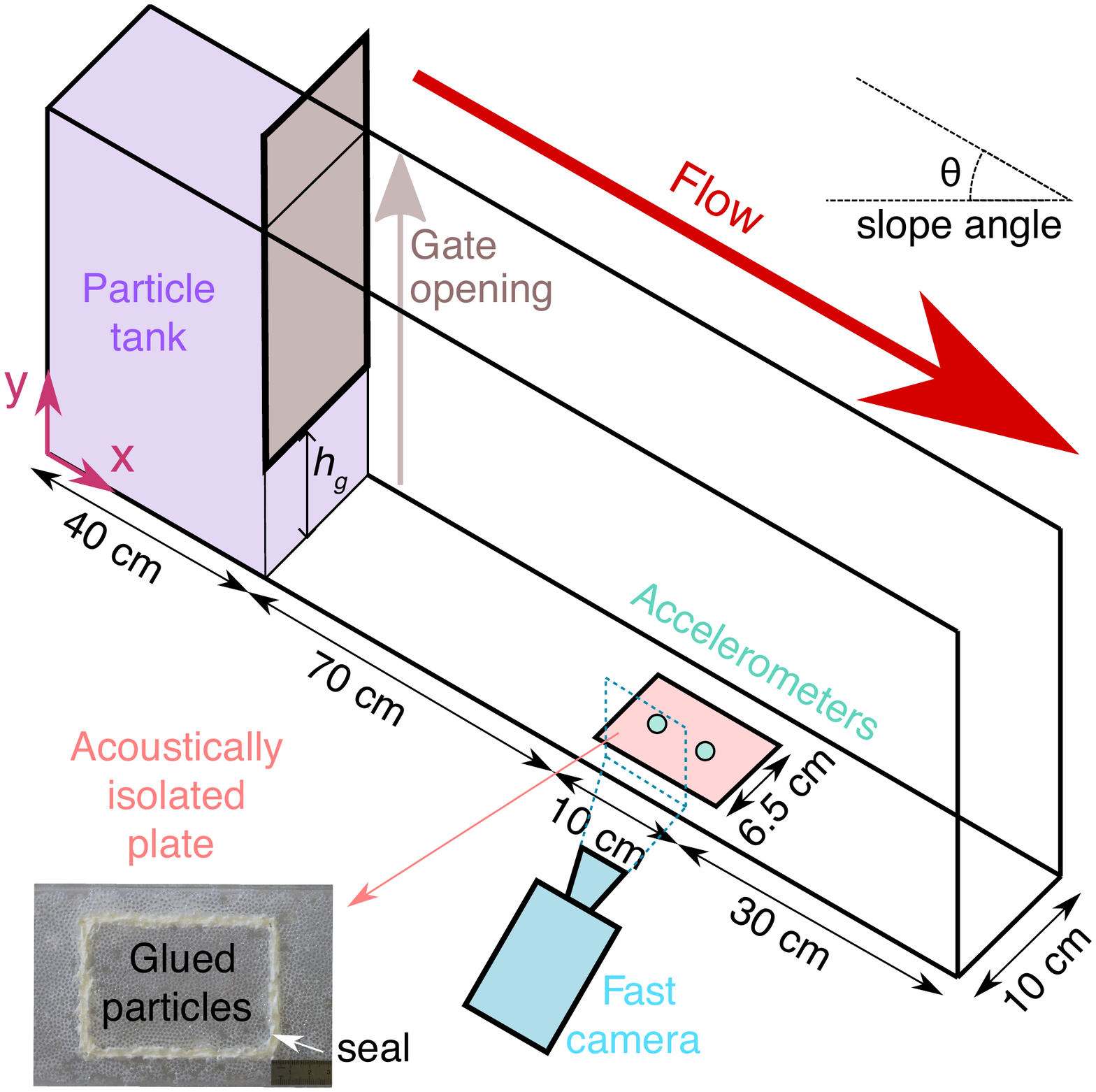}
	\caption{Set-up composed of a narrow inclined channel in which granular flows are created by opening the gate of the upstream tank that contain glass particles. The same particles are glued to the bottom plate to obtain a rough surface. The flow properties are measured by a fast camera and the generated acoustic waves by accelerometers fixed on the channel bottom.}
	\label{fig:set_up}
\end{figure}

The experimental set-up consists of a 1.5~m long chute made of poly(methyl methacrylate) (PMMA), inclined at an angle $\theta$ to the horizontal, with rigid side walls 10~cm apart.
Granular flows are initiated by opening a gate that releases glass particles of diameter $d=2$ mm and density $\rho=2500$ kg~m$^{-3}$, initially stored in a tank (Fig. \ref{fig:set_up}). The resulting flow thickness is related but not equal to the height of the gate that varies between $h_g=4.4$~cm and $h_g=8.5$~cm.
The rough bed is made of the same glass particles glued on the bottom of the PMMA plate with phenyl salicylate, a crystalline substance with low melting point.
As opposed to tape, it prevents the glued particles from vibrating and significantly disturbing the acoustic signal. The two control parameters are the height of the gate $h_g$ and the slope angle of the channel $\theta$ that varies between $\theta=16.5^{\circ}$ and $\theta=18.1^{\circ}$. In this range of inclination angles, almost steady and uniform flows can be observed at about $70$~cm from the gate as discussed below. The characteristics of these flows are summarized in Table \ref{tab:table_parameters}.
At this position, a Photron SA5$^{\mbox{\scriptsize{\textregistered}}}$ fast camera ($5000$ frames per second) records the flow during $2\,\mathrm{s}$ with a field of view of around $50\,\mathrm{mm}$ by $50\,\mathrm{mm}$.
Simultaneously, two accelerometers (bandwidth 10~Hz-54~kHz) record the radiated acoustic waves. These accelerometers are glued, using the same phenyl salicylate as for the particles of the rough surface, on the back of a $10\,\mathrm{cm} \times 6.4\,\mathrm{cm}$ plate isolated acoustically from the rest of the channel bottom.
To isolate the plate, we fixed it to the channel bottom with a silicone sealant (see bottom of Fig. \ref{fig:set_up}).

\begin{table}
  \caption{Parameters of the quasi-steady and quasi-uniform flows obtained in our 9 experiments (referred by the index 1-9): slope angle of the channel $\theta$, thickness of the flow $h$, downslope velocity of the surface particles $V_{xs}$, average downslope velocity $\left<V_x\right>$, average shear rate $\left<\dot{\gamma}\right>$ and average inertial number $<I>$. Note that here $\sqrt{gd}\simeq 0.14$ m/s and $\sqrt{d/g} \simeq 0.014$ s.}
  \centering
  \begin{tabular}{ccccccc}
  \hline
  Index & $\theta\,[^{\circ}] \, (\pm 0.1)$ & $h/d \, (\pm 0.5)$ & $V_{xs}/\sqrt{gd} \, (\pm 0.05)$ & $\left<V_x\right>/\sqrt{gd} \, (\pm 0.05)$ & $\sqrt{d/g} \, \left<\dot{\gamma}\right> \, (\pm 0.01)$ & $\left<I\right> \, (\pm 0.003)$ \\
  \hline
    $1$ & $16.5$ & $17.5$ & $2.15$ & $0.65$ & $0.12$ & $0.070$ \\
    $2$ & $16.5$ & $18.0$ & $2.05$ & $0.55$ & $0.10$ & $0.054$ \\
    $3$ & $16.5$ & $20.0$ & $2.35$ & $0.80$ & $0.12$ & $0.061$ \\
    $4$ & $17.2$ & $15.5$ & $2.50$ & $0.75$ & $0.15$ & $0.094$ \\
    $5$ & $17.2$ & $16.5$ & $2.85$ & $0.90$ & $0.16$ & $0.094$ \\
    $6$ & $17.2$ & $16.5$ & $2.95$ & $1.00$ & $0.17$ & $0.103$ \\
    $7$ & $18.1$ & $14.5$ & $2.02$ & $0.50$ & $0.11$ & $0.074$ \\
    $8$ & $18.1$ & $15.0$ & $2.95$ & $0.90$ & $0.18$ & $0.103$ \\
    $9$ & $18.1$ & $16.5$ & $3.45$ & $1.10$ & $0.21$ & $0.131$ \\
  \hline
  \end{tabular}
  \label{tab:table_parameters}
\end{table}

\section{Optical and Acoustic Methods}

Our objective is to obtain deep quantitative insights into the mean properties of the flow and into its fluctuations and heterogeneity, in order to further interpret the generated acoustic signal in terms of grain scale and mean flow dynamics. Before analysis of these measurements in section \ref{flowchar}, let us detail below the optical and acoustic methods used here to measure flow and acoustic characteristics, respectively. To illustrate the methods, we focus in this section on the two 'extreme' cases: experiment $1$ at $\theta=16.5^{\circ}$ with flow thickness $h=3.5\,\mathrm{cm}$ and surface velocity $V_{xs}=0.30\,\mathrm{m~s^{-1}}$ or experiment 2, and experiment $9$ at $\theta=18.1^{\circ}$ and with $h=3.3\,\mathrm{cm}$ and $V_{xs}=0.48\,\mathrm{m~s^{-1}}$ (Table \ref{tab:table_parameters}).

\subsection{Flow Measurement using Optical Methods}

The flows in all our experiments almost reach a steady and uniform regime: their heights vary by only one particle diameter in space and time (see Fig. \ref{fig:Height_flows} in the Appendix). However, the average heights decrease by around half a particle diameter in the flow direction between $x=0$ and $x=25d$ (Fig. \ref{fig:Height_flows}c in the Appendix), corresponding to an angle of $1^{\circ}$. As a result, our flows are not fully steady as discussed in section \ref{meanflow}. They would have stopped if the channel had been long enough.

\subsubsection{Mean Velocity and Fluctuations}

\begin{figure}
\centering
\includegraphics[width = 12cm]{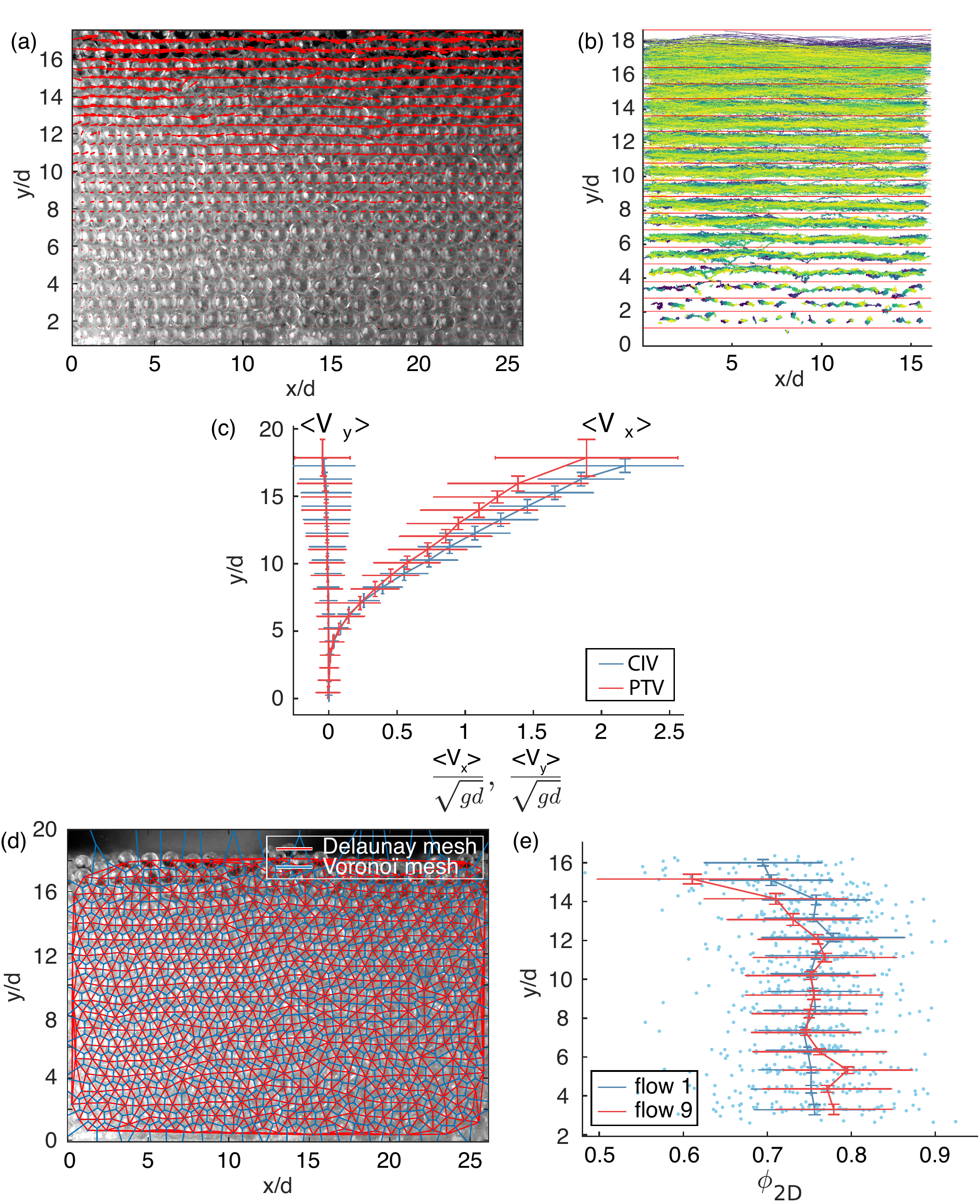}
	\caption{(a) Example (experiment 2) of the velocity field calculated by CIV (red arrows) and (b) superposition of particle trajectories obtained with PTV during $2\,\mathrm{s}$. The organization of the flow into a superposition of layers is clearly visible. In (b), red lines indicate the separation between layers. (c) Example (experiment 2) of mean downslope $\left<V_x\right>$ and normal $\left<V_y\right>$ velocity profiles as a function of the position above the bottom $y$. The associated velocity fluctuations are represented by the horizontal error bars. Vertical error bars correspond to the thickness of the layer within which the velocity has been averaged. Comparison between the measurements made by CIV (blue line) and PTV (red line). (d) and (e) Surface packing fraction of the particles in contact with the lateral wall: (d) manual picking of the particles of flow 1 ($\theta = 16.5^{\circ}$, $h/d = 17.5$) at one instant and (e) deduction of the surface packing fraction (blue dot) per Vorono\"i cell. The average values are plotted in the full blue line. For comparison, the average surface packing fractions of flow 9 ($\theta = 18.1^{\circ}$, $h/d = 16.5$) are plotted in the full red line.}
	\label{fig:particle_tracking}
\end{figure}

We measured particle velocities ${\mathbf{V}}=\left(V_x, V_y\right)$ by Correlation Image Velocimetry (CIV) and Particle Tracking Velocimetry (PTV). CIV divides the images from the movie into boxes and calculates the average displacement into each box by correlation of the graymap between successive images (Fig. \ref{fig:particle_tracking}a). The size of the boxes is a crucial parameter. Boxes too large miss individual particles whereas boxes too narrow do not allow good correlations. Similarly to \citet{Gollin2015a}, the size of the boxes was chosen to be equal to $1.14$ particles. The overlap between boxes is $75\%$. We used the code developed by \citet{Thielicke2014}.

On the other hand, PTV detects and follows the particle positions, making it possible to record their trajectories (Fig. \ref{fig:particle_tracking}b). The particles are semi-transparent and cause complex reflection effects. Consequently, a compromise must be made between the number and relevance of detections. PTV shows that particles are essentially organized into layers that do not really mix during the flow. Mean velocities  $\left<\mathbf{V}\right>=(\left<V_x\right>,\left<V_y\right>)$ are therefore calculated by averaging the measurements within each layer (over 1 particle diameter in the $y$-direction), the borders of which are clearly visible on the PTV images (Fig. \ref{fig:particle_tracking}b). As done for calculating the mean thickness, the averaging is performed over about $16$ particles in space in the downslope direction and over the whole experiment duration ($2\,\mathrm{s}$).

Velocity fluctuations $\delta V$ are computed over the same intervals ($2 \, \mathrm{s}$, $16$ particles in the $x$-direction and $1$ particle in the $y$-direction) by taking the standard deviation of the norm of the velocities:
\begin{equation}
\label{eq:deltaV}
\delta V = \sqrt{\delta {V_x}^2 + \delta {V_y}^2}
\end{equation}
where $\delta {V_i}^2 = \left<(V_i - \left<V_i\right>)^2\right>$ the variance of the velocity along the $i$-direction, with $i=x, y$. For granular systems, the measurement of velocity fluctuations may lead to scale dependency effects due to gradients developing in the flow (see e.g. \citet{Artoni2015b}). Indeed, the thickness $w$ of the layers within which the velocity fluctuations are calculated affects the estimates. Following \citet{Glasser2001}, we showed that the size dependency starts for $w>2d$ (see Fig. \ref{fig:deltaV_window} of \ref{windowsize}). In the following, we will consider velocity fluctuations calculated with a window size $w=d$. Note that when velocity fluctuations are calculated over a smaller time window (e.g. $w=0.2d$), the layering of the flow clearly appears and resembles what was observed by \citet{Weinhart2013} (Fig. \ref{fig:deltaV_window}, \ref{windowsize}). Note that velocity fluctuations of about $0.1\sqrt{gd}$ are measured near the bottom where the mean velocity is zero. This gives an order of magnitude of the error in the measurement of velocity fluctuations ($\sim 0.01$~m~s$^{-1}$).

The profiles of the mean velocity in the downslope $\left<V_x\right>$ and normal $\left<V_y\right>$ directions obtained using CIV and PTV only differ by maximum $10\,\%$ as illustrated in Fig. \ref{fig:particle_tracking}c. In contrast, velocity fluctuations may differ by up to a factor of two between the two methods. This discrepancy has also been observed by \citet{Gollin2015b} and \citet{Gollin2017} and seems to be due to the average nature of CIV,  which is therefore less relevant to measure fluctuations. As a result, PTV measurements will be used in the following, as in \citet{Pouliquen2004}, except for mapping of the spatio-temporal distribution of velocity fluctuations (Fig. \ref{fig:map_deltaV}).

\subsubsection{Packing Volume Fraction}
\label{packing-fraction}

The set-up can only measure the surface packing fraction $\phi_{2D}$ at the lateral walls (Fig.~\ref{fig:particle_tracking}de). The complex light reflections makes it impossible to extract the volume packing fraction $\phi_{3D}$ from $\phi_{2D}$ as proposed by \citet{Sarno2016}. Nevertheless, as is typically observed, we measure an almost constant packing fraction within the flow and a decrease when approaching the free surface (Fig.~\ref{fig:particle_tracking}e). Due to the strong uncertainty in our measurements, the change of $\phi_{2D}$ when increasing the slope angle (i.e. when the inertial number changes) is hard to capture, even though a decrease of $\phi_{2D}$ with increasing inertial number is visible near the surface, in agreement with the literature \citep{Midi2004}. Calculation of the volume fraction shows the layering of the granular flows observed for example in \citet{Artoni2015b} and \citet{Weinhart2013}.

\subsubsection{Frequency of Particle Oscillations}
\label{particle-oscillations}

\begin{figure}
\centering
\includegraphics[width = 8cm]{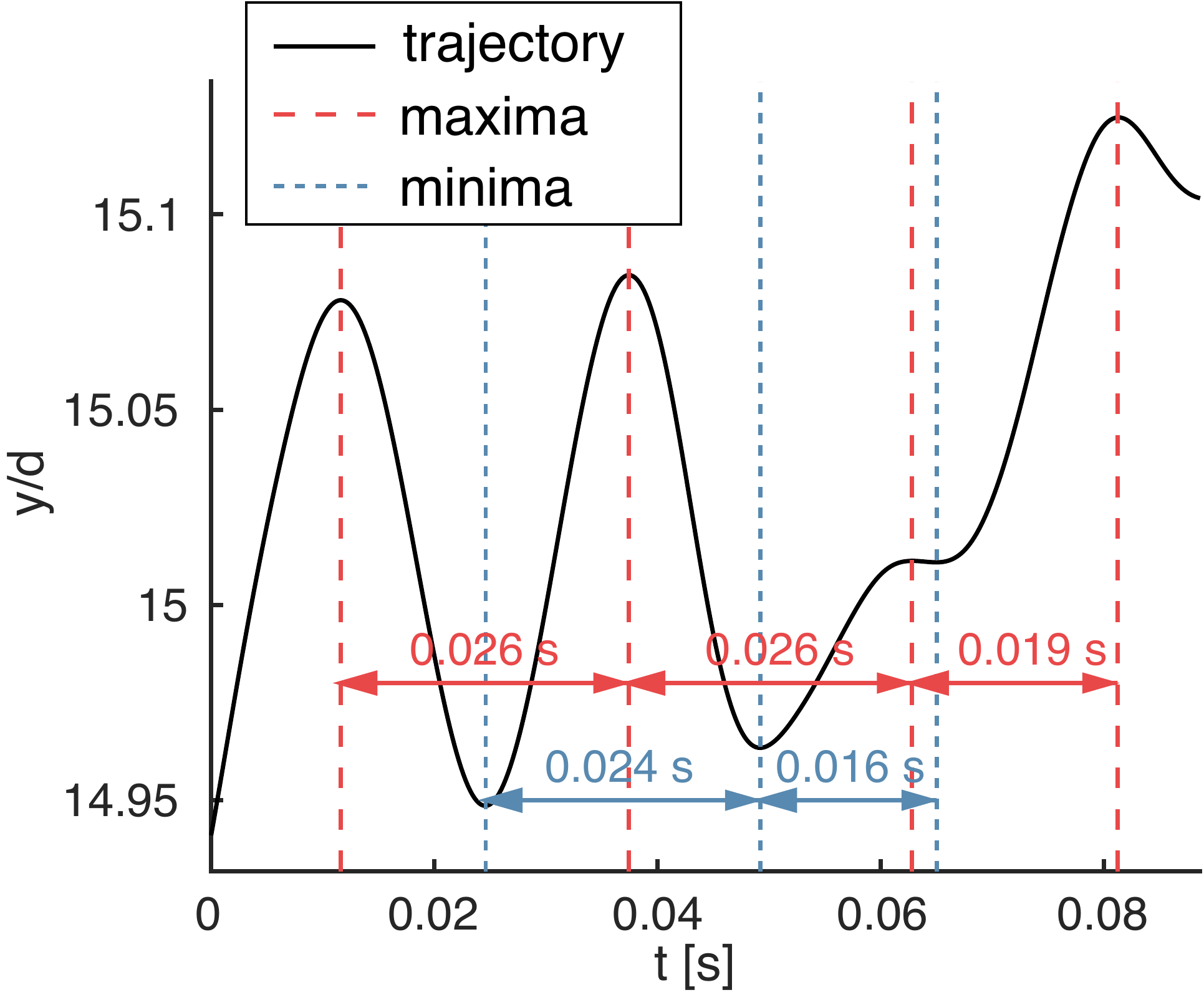}
	\caption{Example (experiment 2) of vertical particle oscillations captured by PTV for a particle located at the surface of the flow:
The smoothed trajectory shows the average period of the oscillations $T\simeq 0.02$~s.}
	\label{fig:fy_particles}
\end{figure}

During the flow, vertical oscillations of the particles can be observed related to compression/dilatation effects occurring when one layer passes over another (see Movies 1 and 2 in supplementary material). The frequency of these oscillations can be captured in the PTV measurements of the trajectory of the particles
located at the surface (Fig. \ref{fig:fy_particles}). Indeed, several oscillations can be observed before loosing the particle tracking owing to the relatively high velocity of these particles. On the contrary, for the particles located deeper in the flow, the oscillations generally occur when the particle tracking has already been lost.
The oscillation frequency is calculated by filtering the particle trajectory with two filters and taking the median of $1/T$, where $T\simeq 0.02$~s is the time separating
successive maxima and minima of the trajectory (Fig. \ref{fig:fy_particles}). More precisely, the first filter is a normalized median filter adapted
from \citet{Westerweel2005} for vectors applied on each trajectory component (neighborhood radius of $5$ successive positions, noise threshold level
of $0.10$ and threshold of $1$), and the second filter is a second order zero-phase low pass filter (cut-off frequency of $50 \, \mathrm{Hz}$).
The median filter has been chosen to suppress random fluctuations.

\subsection{Elastic Wave Measurements}

The elastic waves generated by the granular flows and by their interactions with the bottom are recorded by two accelerometers glued to the isolated plate (Fig. \ref{fig:acoustics}a). It is assumed here that the accelerometers mainly record the vibrations generated by the section of granular flow over the plate. Isolation of the plate from the rest of the flume was verified by comparing the signals recorded by accelerometers glued to these two elements. 

\subsubsection{Radiated Elastic Power}

The radiated elastic power over duration $\Delta t$ is $\Pi_{el}=W_{el}/\Delta t$, where $W_{el}$ is the radiated elastic energy. The acoustically isolated plate is small compared to the characteristic viscoelastic attenuation length of energy in PMMA. As a result, the waves are reflected many times at the boundaries of the plate leading to a diffuse elastic field, i.e. homogeneously distributed over the plate and equipartitioned. The elastic energy can then be computed using the diffuse field theory proposed by \citet{Farin2016}:
\begin{equation}
\label{eq:Wel_diffus}
W_{el} = M \, \gamma \, v_g \, \times \int_{\Delta t} v_z^2(t)\mathrm{d}t
\end{equation}
where $M \approx 80\,\mathrm{g}$ is the mass of the isolated piece of plate, $\gamma \approx 3\,\mathrm{m^{-1}}$ its average viscoelastic attenuation and $v_g \approx 1000$ m~s$^{-1}$ the average group velocity of the radiated acoustic waves ($A_0$ Lamb waves). The value of $\gamma$ is obtained by measuring the response of the plate at various distances with a source and a vibrometer and the value of $v_g$ by calculating the dispersion relation of the $A_0$ Lamb modes of the plate following \citet{Royer2000}.
 A large time window $\Delta t =0.2\,\mathrm{s}$ is selected in order to consider only slow changes of $\Pi_{el}$, with error bars representing the standard deviation. The fast fluctuations will be characterized in the next section. An example of radiated elastic power computation is presented in Fig. \ref{fig:acoustics}a.

\subsubsection{Frequency Content}
\label{par:freq}

\begin{figure}
\centering
\includegraphics[width = 10cm]{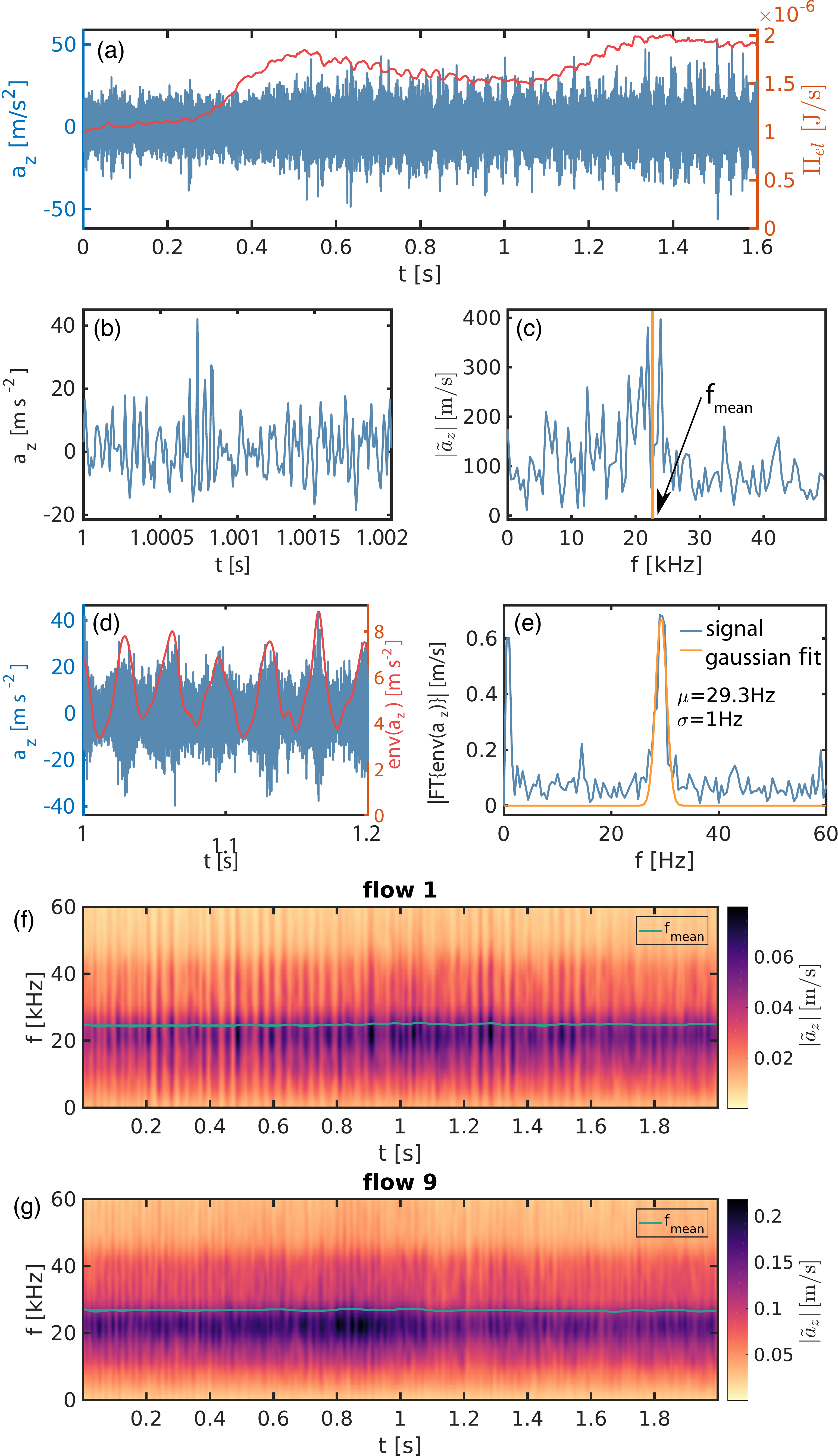}
	\caption{Acoustic signal of the flow number $2$: (a) acceleration of the vibration (blue) and associated elastic power (red), (b) enlargement of the acoustic signal and (c) associated spectra, (d) envelope (red) of the acoustic signal (blue) and (e) associated spectra of the envelope. (f) and (g) Spectrogram of the signal of (f) experiment $1$ ($\theta=16.5^{\circ}$, $h=3.5\,\mathrm{cm}$, $V_{xs}=0.30\,\mathrm{m~s^{-1}}$) and (g) experiment $9$ ($\theta=18.1^{\circ}$, $h=3.3\,\mathrm{cm}$, $V_{xs}=0.48\,\mathrm{m~s^{-1}}$). The dark colors show the mean frequency (a few tens of kHz) with its mean value represented by a light blue horizontal line, whereas the so-called modulation frequency, 1000 times smaller, is related to the distance between the vertical stripes (a few tens of Hz).}
	\label{fig:acoustics}
\end{figure}

The spectrogram of the acoustic signal shows a mean frequency of around $20-30\,\mathrm{kHz}$, corresponding to the average height of the dark bands
in Fig. \ref{fig:acoustics}f,g, that slightly increases when the slope angle increases. The mean frequency is calculated as follows:
\begin{equation}
\label{eq:fmean_exp}
f_{mean} = \frac{\int_0^\infty |\tilde{A}_z(f)| f \mathrm{d}f}{\int_0^\infty |\tilde{A}_z(f)| \mathrm{d}f}
\end{equation}
within time windows of $2\,\mathrm{ms}$ (Fig. \ref{fig:acoustics}b,c), where $\tilde{A}_z$ is the Fourier transform of the acceleration $a_z(t)$ of the vertical vibration.
Amplitude spectra are not studied beyond $54 \, \mathrm{kHz}$ because of the limit of the accelerometer responses (signal polluted by the accelerometer resonances).

Vertical stripes can be identified on the spectrograms (Fig. \ref{fig:acoustics}f,g). The distance between these stripes decreases as the slope angle increases.
This corresponds to a so-called modulation amplitude frequency of around $25-50\,\mathrm{Hz}$, about $500$ to $1000$ times smaller than the mean frequency defined in Eq.~(\ref{eq:fmean_exp}). To calculate the modulation frequency, we first extract the envelope of the signal (absolute value of the signal) and apply a low pass filter (cut-off frequency empirically fixed at $75\,\mathrm{Hz}$). Then, the mean oscillation frequency is determined by fitting a Gaussian in the Fourier space (Fig. \ref{fig:acoustics}d,e).

\section{Flow Characteristics} \label{flowchar}

Our objective here is to capture the relationship between mean flow properties and fluctuations that are expected to play a role in acoustic emissions. Note that the flow measurements are made at the side walls. It is well known in regard of the bulk that the wall boundaries significantly affect the mean flow quantities and their fluctuations, as will be discussed below (see e.g. \citet{Taberlet2003, Jop2005, Jop2007, Artoni2015, Mandal2017, Fernandez2018}).

\subsection{Mean Flow} \label{meanflow}
\begin{figure}
\centering
\includegraphics[width = \linewidth]{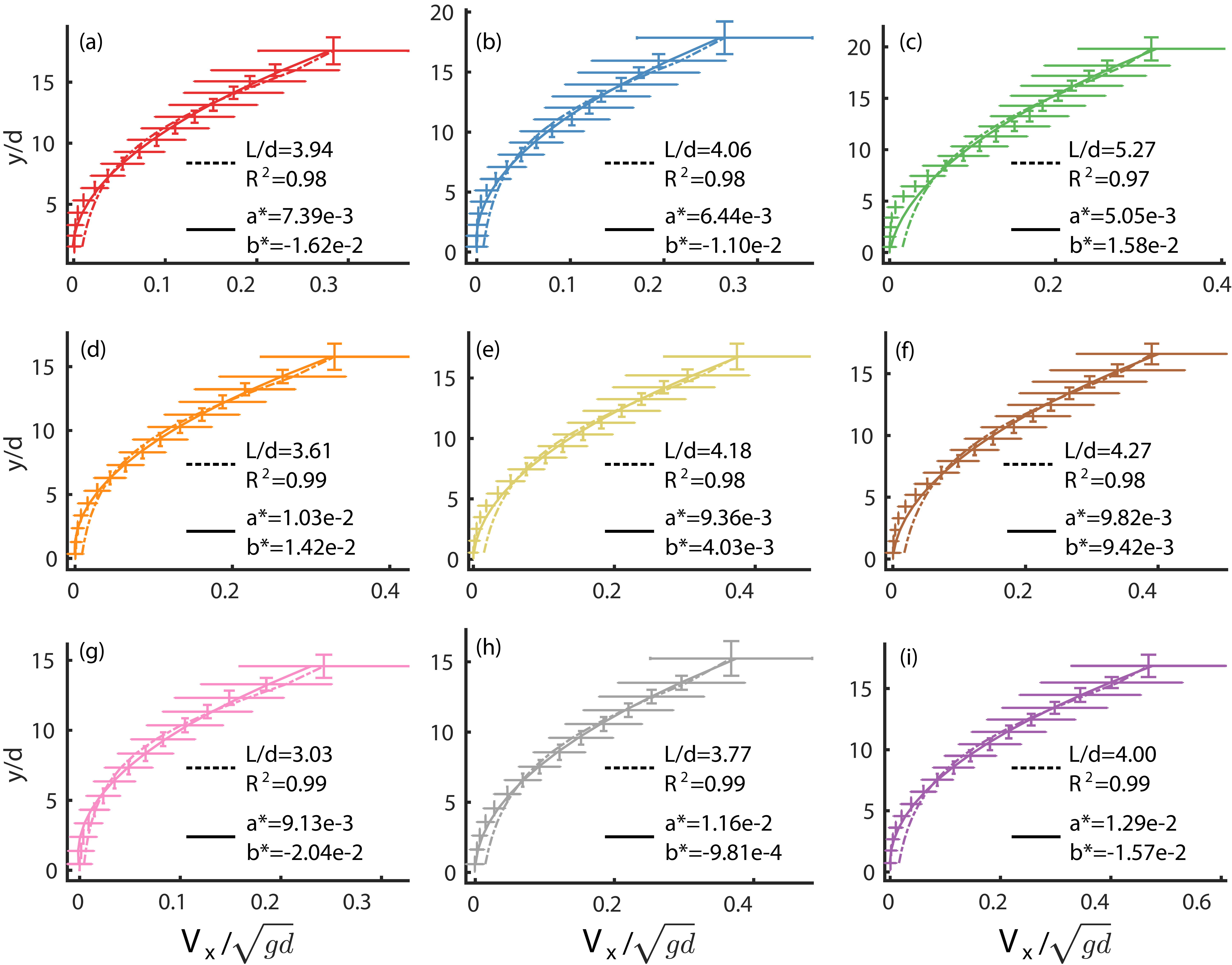}
	\caption{Velocity profiles of all the experiments with letters (a) to (i) referring to flows 1 to 9, corresponding to the angles (a-c) $\theta=16.5^{\circ}$, (d-f) $\theta=17.2^{\circ}$ and (g-i)
$\theta=18.1^{\circ}$, each associated with increasing flow thicknesses (see Table \ref{tab:table_parameters} for details).
Two theoretical profiles have been fitted: the ones given by Eq. (\ref{eq:Josserand}) in dashed lines and a 2nd order polynomial
($V_x/\sqrt{gd}=a^* {\left(y/d\right)}^2 + b^*(y/d)$) in full lines. For all polynomial fits, $R^2 \geq 0.99$.}
	\label{fig:velocities_all}
\end{figure}

The nearly uniform and steady flows confined in a narrow channel inclined at slope angles between $16.5^\circ$ and $18.1^\circ$ obtained here are similar to those observed by \citet{Hanes2000} in similar settings. In these flows, the downslope velocity $V_x(y)$ is maximum at the free surface, decreasing down to zero near the bottom (Fig. \ref{fig:velocities_all}). Such convex velocity profiles are observed in flows confined in narrow channels (see e.g. \citet{Ancey2001,Courrech2003, Taberlet2003, Midi2004, Jop2005, Jop2007, Mandal2017}) and differ from Bagnold-like velocity profiles obtained for steady and uniform flows in wide channels (see \cite{Midi2004} or Fig. 4 of \citet{Fernandez2018}). These profiles have a shape that can be approximately fitted by the velocity profiles assumed in \citet{Josserand2004} to describe heap flows:
\begin{equation}
\label{eq:Josserand}
1 - \frac{V_x(y')}{V_x(y'=0)} = {\left(\frac{1-e^{-y'/Y}}{1+(\frac{\phi_M}{\phi_m}-1)e^{-y'/Y}}\right)}^{3/2}
\end{equation}
where $y' = y_h - y$ and $y_h$ is the height of the flow surface, $Y$ a fitting parameter, $\phi_M=0.65$ and $\phi_m=0.5$ is the loose and dense random packing fraction,
respectively. Fig. \ref{fig:velocities_all} shows that Eq. (\ref{eq:Josserand}) fits our experimental data quite well except near the bottom
for experiments with thick flow depth $h$, where the horizontal velocity is non-zero at the base. Second order polynomials
($V_x/\sqrt{gd}=a^* {\left(y/d\right)}^2 + b^*(y/d)$) give even better results, especially near the bottom.
We will therefore use these polynomial fits to calculate the shear strain rate $\dot{\gamma}=\partial V_x / \partial y$.

\begin{figure}
\centering
\includegraphics[width=12cm]{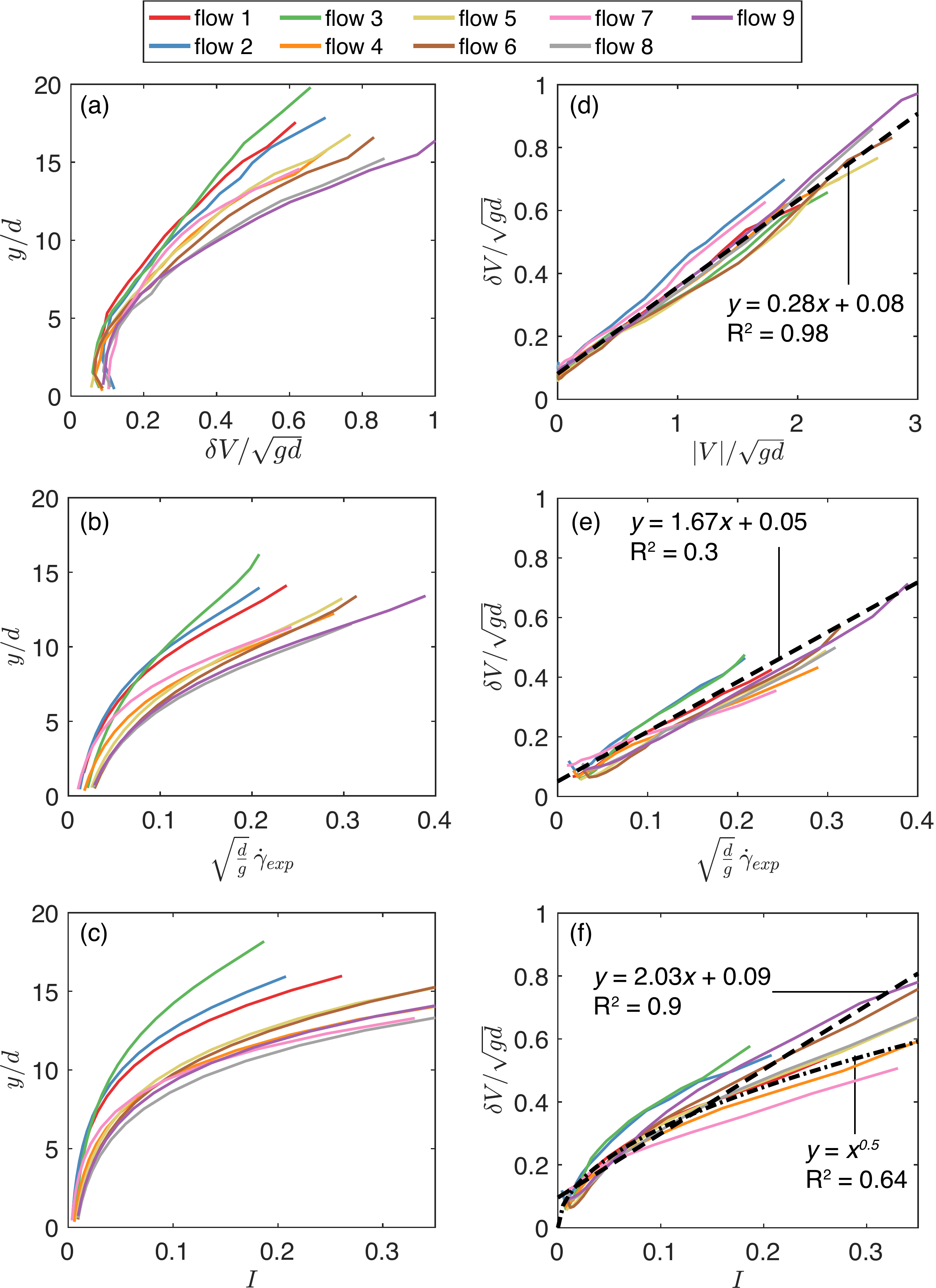}
	\caption{(a) Normalized fluctuating speed $\delta V/\sqrt{gd}$, (b) Normalized shear rate $\sqrt{d/g}\dot{\gamma}$ and (c) Inertial number $I$, computed using the second order polynomials that provide the best fit, as a function of flow depth $y/d$, for all of the experiments (colors). (d) to (f) Normalized fluctuating speed $\delta V/\sqrt{gd}$ as a function of (d) the flow normalized average speed $|V|/\sqrt{gd}$, (e) the normalized shear rate $\sqrt{d/g}\dot{\gamma}$ and (f) inertial number $I$. In panels (d) to (f), dashed lines show fits of the data with linear laws. In panel (f), the dash-dotted line shows a power-law (square root) fit of the data.}
	\label{fig:deltaV_gamma_I}
\end{figure}

The shear strain rate $\dot{\gamma}$ decreases from the surface down to the bottom (Fig. \ref{fig:deltaV_gamma_I}b).
Granular flows are characterized by the inertial number $I=\dot{\gamma}d/\sqrt{P/\rho_s}$, where $P$ is the pressure
taken here as hydrostatic ($P=\rho_s \phi g \cos(\theta) (h-y)$):
\begin{equation}
I(y) = \frac{\dot{\gamma}(y)d}{\sqrt{\phi g \cos(\theta) (h-y)}}
\label{eq_Inertial_number}
\end{equation}
The packing fraction is approximated by $\phi=0.6$ \citep{Jop2005} because we do not have access to the packing fraction in the bulk of the flow
(see section \ref{packing-fraction}). As the velocity profiles are not Bagnold-like, the inertial number is not constant with depth here,
but decreases from the surface to the bottom (Fig. \ref{fig:deltaV_gamma_I}c).

\subsection{Velocity Fluctuations}
\label{fluct-heterogen}

The high-frequency acoustic signal generated by granular flows is expected to be mainly due to particle collisions,
even though friction may play a role that will not be considered here \citep{Huang2007, Michlmayr2013}.
Such collisions occur when neighboring particles have different velocities, in particular related to velocity fluctuations.
Velocity fluctuations, or their squared values called granular temperature \citep{Goldhirsch2008},
\begin{equation}
T=\left<\delta V^2 \right>
\end{equation}
where $\left<.\right>$ means the average over volume and time ranges, are known to be significant in granular flows.
Granular temperature is however generally not accounted for explicitly in the rheology of dense granular flows,
except in the extended kinetic theory \citep[e.g.][]{Berzi2014,Gollin2017}.
Indeed, the relationship between velocity fluctuations and the inertial number or other mean flow quantities has not yet been thoroughly investigated in dense granular flows. Indeed, they are difficult to measure experimentally, and even more in the field \citep{Berzi2011,Hill2014}.
The acoustic power, much easier to measure, may provide a unique tool to obtain quantitative measurements of granular temperature as will be investigated below.

Fig. \ref{fig:deltaV_gamma_I}a shows that velocity fluctuations decrease from the surface to the bottom for all experiments and increase with slope angle.
Using discrete element modeling, \citet{Hanes2000} showed that the granular temperature profile is very different at the side wall than within the core of the flow: the granular temperature at the surface is the same at the side walls and across the flow but it increases with depth in the middle of the flow, while it decreases at the side walls, as observed here.

Even though $\delta V$ looks regular when averaged over volume and time, Figs. \ref{fig:map_deltaV}(a) and (b) in the Appendix and Movies 3 and 4 in the supplementary material illustrate the existence of transient vortices of velocity fluctuations in our experiments, as observed by \citet{Kharel2017}.
The size and intensity of these transient vortices seem to be related to the flow regime, leading to strong variations
of velocity fluctuations (in space and time) when approaching jamming, and thus possibly contributing to generate acoustic emissions in these regions.
The correlation length of these velocity fluctuations is around 1 grain diameter in the $y$-direction and can reach up to 8~d in the $x$-direction,
decreasing with increasing slope (see Fig. \ref{fig:Correlations_all} in \ref{correlation}).

\subsection{Relationship Between Mean Properties and Fluctuations}

Granular temperature is expected to scale with the square of the shear strain rate so that $\delta V \propto \dot{\gamma}$ \citep[see e.g.][]{Pouliquen2004, Andreotti2013}.
Such a linear relationship between $\delta V$ and $\dot{\gamma}$ seems indeed to be satisfied (Fig. \ref{fig:deltaV_gamma_I}e), in very good agreement
with what was found at the surface of granular flows by \cite{Pouliquen2004} or in other configurations \citep{Midi2004}.
If we try to fit the data by a power law, we get a power equal to 2 with high $R^2$. 
A higher $R^2$ is found when trying to relate the velocity fluctuations to the mean downslope velocity $\left<V_x\right>$ (Fig. \ref{fig:deltaV_gamma_I}d).
The slightly higher $R^2$ obtained when relating velocity fluctuations to the mean velocity compared to the strain rate may result from errors related
to the calculation of the gradient of the measured velocity profile.
The power law between velocity fluctuations and the inertial number is less clear, with a smaller $R^2$ (Fig. \ref{fig:deltaV_gamma_I}f). This is also possibly due to the errors in the calculation of $I$. As a result, velocity fluctuations averaged in time and along one layer of grains scale very well with shear rate and with mean velocity and to a lesser extent with the inertial number:
\begin{equation}
\delta V \propto <V_x> \propto \dot{\gamma} \propto I^{0.5}.
\end{equation}

\section{Signature of Flow Dynamics in the Acoustic Signal}
Our objective is to quantitatively relate the characteristics of both the seismic signal and the flow to
(i) get physical insights into the sources of acoustic emission and (ii) propose empirical scaling laws that can be used to recover flow properties
from the recorded acoustic waves. As the range of configurations (slope angle, thickness) investigated here is not very large,
it is hard to discriminate between power laws or linear trends. We will therefore systematically test these two types of empirical fits
and quantify the associated $R^2$.

\subsection{Acoustic Frequencies}

\subsubsection{Order of Magnitude of Expected Frequencies}
\label{section_discussion_freq}

Let us first discuss the order of magnitude of the expected frequencies in the measured acoustic signal generated by the granular flows and their potential causes, based on our setup and on the observation of flow dynamics described in the previous sections.

The main frequency of the signal is expected to be caused by particle collisions between two spheres of diameter $d$ at relative velocity $\delta V$ and to approximately scale with the inverse of the contact time calculated by the Hertz contact theory \citep{Farin2015}.
For impacts on smooth plates, 
this leads to the following expression for the main frequency
\begin{equation}
f_{Hertz}=a_0' \, d^{-1} \, \delta V^{1/5},
\label{eq:fmean|hertz}
\end{equation}
where
\begin{equation}
a_0' \approx 0.90 \, {\left(\frac{E^*}{2 \pi \rho}\right)}^{2/5} \approx 140 (\mathrm{I.S.U.})
\label{eq_a0}
\end{equation}
and $E^*$ is the effective Young modulus such that $1/E^*=\left(1-\nu_s^2\right)/E_s+ \left(1-\nu_p^2\right)/E_p$.  $\nu_s=0.2$, $\nu_p=0.37$, $E_s=74 \, \mathrm{GPa}$
and $E_p=4.4 \, \mathrm{GPa}$ are respectively the Poisson's ratios and the Young's moduli of the constitutive materials of the particle (glass) and the impacted plate (PMMA),
and $\rho=2500 \, \mathrm{kg/m^3}$ is the bulk density of the particles.
This leads here to $30 \, \mathrm{kHz} < f_{Hertz} < 48 \, \mathrm{kHz}$ for $0.1 \times \sqrt{gd} <\delta V < \sqrt{gd}$, with $\sqrt{gd}=0.14$ m~s$^{-1}$. However, the frequency of an impact on a
rough bed is less than on a smooth bed as shown in Fig. 9 of \citet{Farin2018}. In their case, involving steel particles, the mean frequency
for impacts on rough beds was about 2/3 to half of the mean frequency over a smooth bed. If we assume similar behavior in our case, we could expect impact frequencies
of about $15 \, \mathrm{kHz} < f_{Hertz} < 30 \, \mathrm{kHz}$. Note that attenuation within the granular media, which is higher for high frequencies, may also decrease the measured mean frequency.

\begin{figure}
\centering
\includegraphics[width= \linewidth]{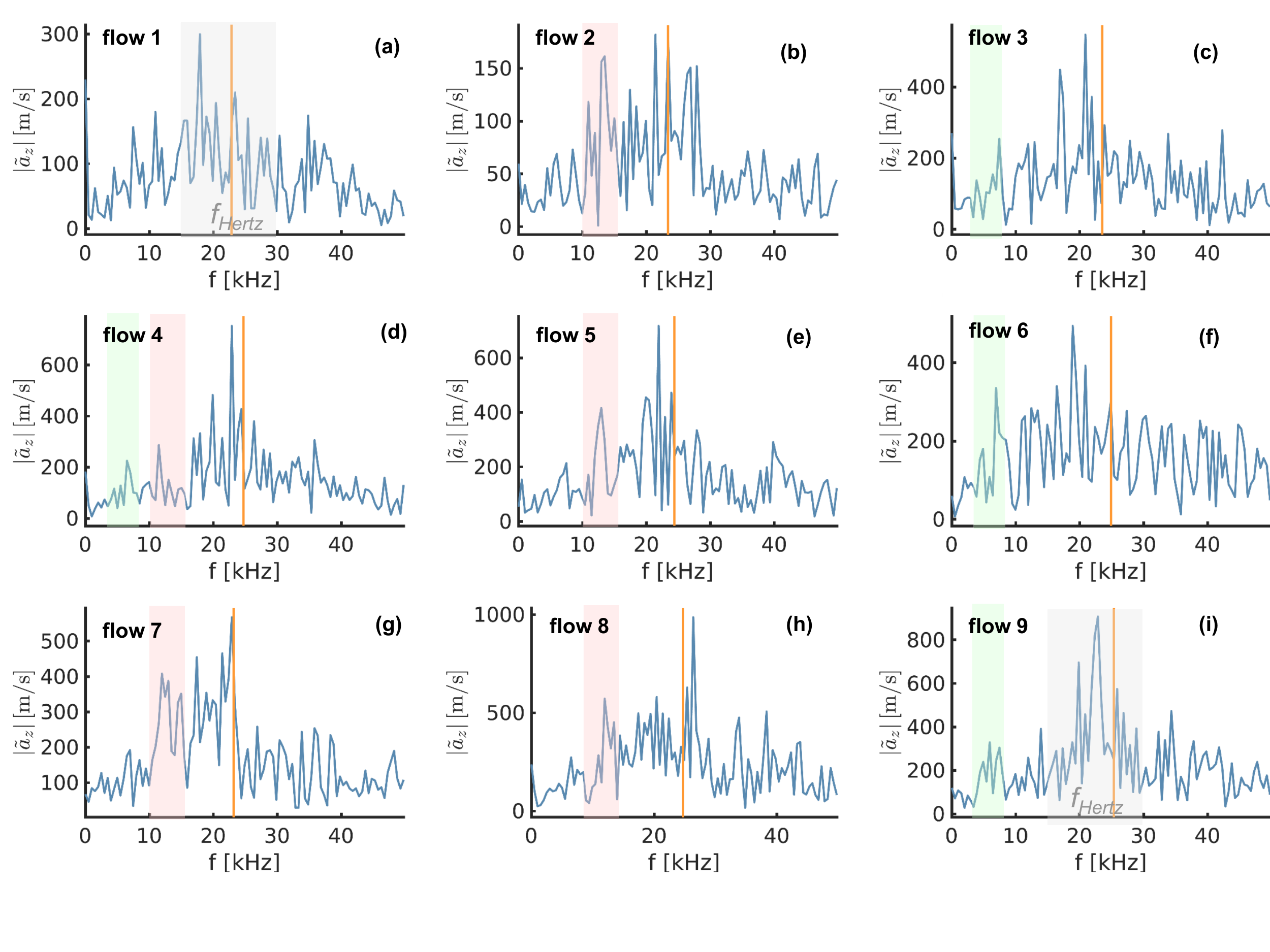}
\caption{High-frequency ($f>1$~kHz) spectral amplitude measured for all flows. Letters (a) to (i) refer to flow numbers 1 to 9 corresponding to angles (a-c) $\theta=16.5^{\circ}$, (d-f) $\theta=17.2^{\circ}$ and (g-i) $\theta=18.1^{\circ}$, each associated with increasing flow thicknesses (see Table \ref{tab:table_parameters} for details).
The orange lines correspond to the mean frequencies. Light gray areas in Fig. (a) and (i) correspond to the range of expected Hertz's frequencies $f_{Hertz}$, light pink areas to the frequency range associated with plate resonance $f_{p1}$ and $f_{p2}$ and light green areas to the frequency range associated with waves trapped in the granular layer $f_h$.}
	\label{fig:f_specters}
\end{figure}

In contrast, the vertical oscillations of the particles due to the motion of one layer over another (see section \ref{particle-oscillations}, Fig.~\ref{fig:acoustics}fg) are a possible cause of the so-called modulation frequencies $33 \, \mathrm{Hz} < f_{mod} < 52 \, \mathrm{Hz}$, about 500 times smaller than $f_{mean}$, shown in Fig.~\ref{fig:f_specters}. This modulation frequency is of the order of magnitude of $\delta V /d$, corresponding to a typical frequency between collision events.

On the other hand, frequencies around $f_{h}\simeq 3-7$ kHz in the signal may originate from the typical period of the acoustic wave front propagation though the flow thickness $h=3$~cm, if we assume an acoustic wave velocity in granular flows of 100-200~m~s$^{-1}$
(see e.g. \citet{Mouraille2008,HostlerPHD2004,Hostler2005}). Note that the velocity of acoustic signals in granular material varies
strongly depending on the confining pressure, packing fraction, material involved, etc. \citet{Liu1993} found values varying from about $60$ to $280\,\mathrm{m~s^{-1}}$
depending upon the kind of velocity measured, \citet{Wildenberg2013} between $80\,\mathrm{m~s^{-1}}$ and $150\,\mathrm{m~s^{-1}}$ and \citet{Bonneau2008} between $40\,\mathrm{m~s^{-1}}$ and $80\,\mathrm{m~s^{-1}}$.

Observations show that the flow thickness oscillates slightly with time (see Fig. \ref{fig:Height_flows} in the Appendix), possibly due
to compression/dilatation waves in the media or to the complex heterogeneity of the flow (see section \ref{fluct-heterogen} and Fig. \ref{fig:map_deltaV} in the Appendix).
The typical period of these oscillations is 1~s, possibly generating waves at frequencies $f_{flow}\simeq 1$~Hz. 

Movies of velocity fluctuations (Movies 3 and 4 in the supplementary material) show the building and destruction of vortices of velocity fluctuations (cf Fig. \ref{fig:map_deltaV} in the Appendix). These vortices may be similar to the turbulent vortices that develop in rivers and apply fluctuating forces on the bed roughness, generating seismic signals over a wide frequency range 1-10$^5$~Hz \citep{Gimbert2014}. Turbulent vortices form close to the flowing static interface due to the shear stress applied by the flow on the bed. The formed vortices enlarge by coalescence until they reach the thickness of the flow, then break up into smaller vortices, transferring flow energy towards the smaller scales \citep{Kolmogorov1941}. The highest frequencies generated by the vortices are related to the minimum vortex size, i.e. the Kolmogorov microscale, which may not be reachable in a granular flow because the minimum vortex scale is in theory at least two particle diameters $2d$. Therefore, in granular flows we expect lower frequencies generated by vortices than those that can be observed in a liquid flow. The typical size of the observed vortices in our granular flows is about 5-8$d\simeq 1-1.6$~cm and they travel within the flow at velocities of around 1~m~s$^{-1}$. Therefore, these granular vortices may generate waves at frequencies $f_v\simeq 60-100$~Hz.

Finally, the resonance of the $10$ cm$\times6.5$ cm acoustically isolated plate gives rise to frequencies $f_{p1}\simeq 15$ kHz and $f_{p2}\simeq 10$ kHz if we assume a wave velocity in the plate of $1000 \, \mathrm{m~s^{-1}}$. Let us now analyze the frequency content of the measured signal and compare it to these expected frequencies.

\subsubsection{Comparison with Measured Frequencies}
\label{compare-frequency}

\begin{figure}
\centering
\includegraphics[width = \linewidth]{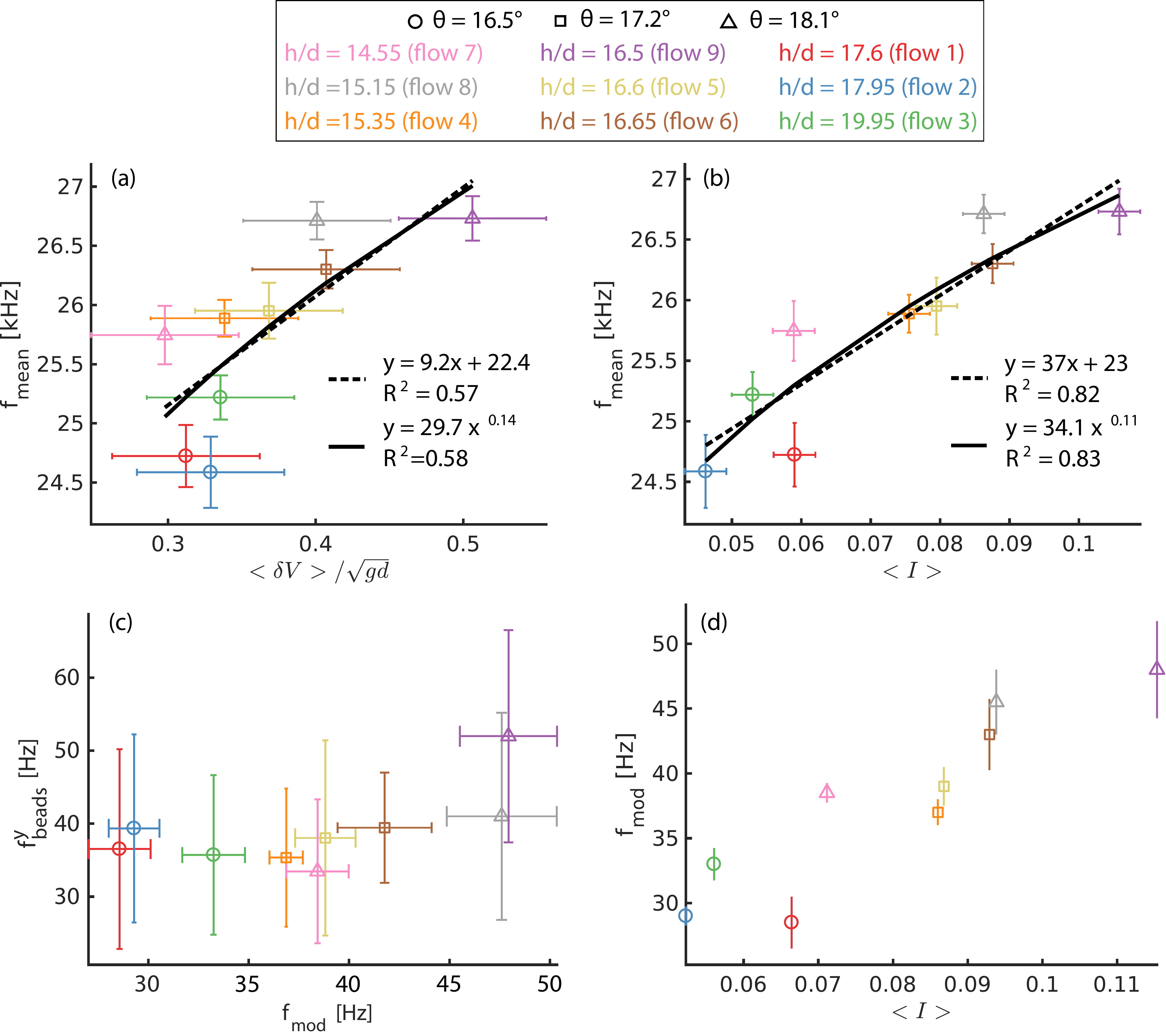}
	\caption{Mean frequency $f_{mean}$ as a function of (a) normalized average fluctuating speed $<\delta V>/\sqrt{gd}$ and (b) the average inertial number $<I>$. (c) Vertical particles oscillation frequency $f^y_{particles}$ as a function of the frequency of the acoustic amplitude modulation. (d) Acoustic modulation frequency $f_{mod}$ as a function of the average inertial number $<I>$.}
	\label{fig:fmean_scalings}
\end{figure}

At high frequencies ($f> 1$ kHz), Fig. \ref{fig:f_specters} shows that the mean frequencies in our experiments are in the range $24.5$ kHz$<f_{mean}<27$ kHz.
This clearly corresponds to the order of magnitude of the frequencies $15$ kHz $<f_{Hertz}< 30$ kHz due to collisions in the Hertz contact theory (frequency range highlighted in light gray in Fig. \ref{fig:f_specters}(a) and \ref{fig:f_specters}(i)). The values of $f_{mean}$ are closer to the maximum expected frequencies, corresponding to higher velocity fluctuations, and thus to the particles located near the free surface.
Fig. \ref{fig:fmean_scalings} shows that the mean frequency $f_{mean}$ increases with the amplitude of the mean velocity fluctuations and mean inertial number. The mean values are obtained by averaging the quantities over the flow depth.
Fitting this increase with affine or power law ($f_{mean} \propto \delta V^{0.14}$) relationships give approximately the same $R^2\simeq 0.58$
(see details in Fig. \ref{fig:fmean_scalings}). The power exponent $0.14$ is not far from the theoretical value predicted by the Hertz theory, i.e. $1/5=0.2$ (see Eq. \ref{eq:fmean|hertz}). Even the coefficient $a_0'/d \simeq 7 \times 10^4$ is not so far from the coefficient of the power law fit $3 \times 10^4*\sqrt{gd}^{0.14}\simeq 2.3 \times 10^4$, supporting the interpretation that these mean frequencies come from the Hertzian contact between particles of relative velocity $\delta V$.  There is a better collapse of the experimental data when 
$f_{mean}$ is represented as a function of $I$, i.e. $f_{mean} \propto I$ or $f_{mean} \propto I^{0.1}$, or as a function of shear strain rate, i.e.
$f_{mean} \propto \dot{\gamma}$ or $f_{mean} \propto \dot{\gamma}^{0.1}$ , leading to $R^2\simeq 0.82$ and $0.83$, respectively. The fit of $f_{mean}$ as a function of $\dot{\gamma}$ gives $R^2\simeq 0.79$.
Even though no clear peaks appear in the high-frequency spectra, some peaks are observed at frequencies $3<f<10$ kHz for almost all the flows, which may correspond to waves trapped within the flowing granular layer $3<f_h<7$ kHz as highlighted for example in light green in Fig. \ref{fig:f_specters}(c), (d), (f), and (i).
Other peaks appear at frequencies between 10 and 20~kHz that may be related to the plate resonance ($f_{p1}\simeq 10$~kHz and $f_{p2}\simeq 15$~kHz) as illustrated
in light pink in Fig. \ref{fig:f_specters}(b), (d), (e), (g), and (h).

\begin{figure}
\centering
\includegraphics[width=\linewidth]{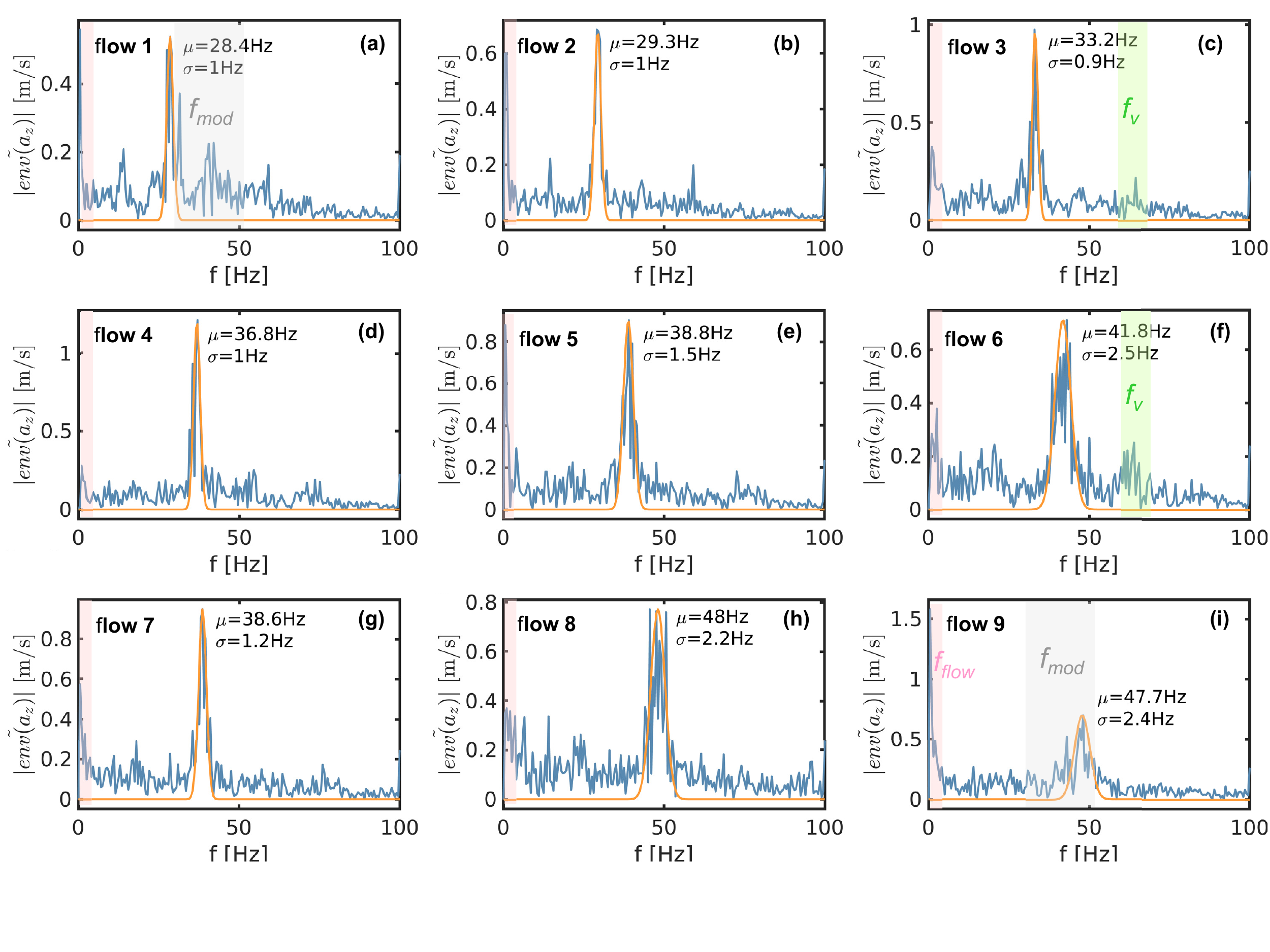}
	\caption{Low-frequency ($f<100$ Hz) spectral amplitude measured for all flows. Letters (a) to (i) refer to flow numbers 1 to 9 corresponding to angles (a-c) $\theta=16.5^{\circ}$, (d-f) $\theta=17.2^{\circ}$ and (g-i)
$\theta=18.1^{\circ}$, each associated with increasing flow thicknesses (see Table \ref{tab:table_parameters} for details).
The orange curves correspond to the Gaussian fits (see Fig. \ref{fig:acoustics}e). Light gray areas in Fig. (a) and (i) correspond to the frequency range associated with particle oscillations $f_{mod}$, light pink zones on all the figures correspond to the frequency range of flow oscillations $f_{flow}$ and light green zones to frequency range of vortices $f_v$.}
	\label{fig:fmod_specters}
\end{figure}

In the low-frequency range, Fig. \ref{fig:fmod_specters} shows clear frequency peaks between 28~Hz and 50~Hz. These frequencies of the acoustic amplitude modulation are clearly in the range of the frequencies $f_{mod}$ associated with the vertical oscillation of the particles at the surface of the flow (Fig. \ref{fig:fmean_scalings}c). Indeed, despite high error bars, they are both between approximately $30$ and $60 \, \mathrm{Hz}$ (frequency range of $f_{mod}$ is highlighted in light gray in Figs. \ref{fig:fmod_specters}(a) and \ref{fig:fmod_specters}(i)).
The acoustic amplitude modulation frequency increases as a function of the inertial number (Fig. \ref{fig:fmean_scalings}d).
Almost all the flows exhibit an increase of spectral amplitude at frequencies between 1 Hz to 3~Hz. This may correspond to the
frequencies of flow oscillations $f_{flow}\simeq 1$~Hz. Some peaks at 15 to 25~Hz also appear for some flows.
Some flows also show a small increase of spectral amplitude of around 60-70~Hz (see Fig. \ref{fig:fmod_specters}(c) and (f)
where this frequency range is highlighted in light green) that could be compatible with frequencies associated with vortices of the velocity fluctuations $f_v\simeq 60-100$~Hz.

\subsection{Acoustic Power}

\subsubsection{Power Laws and Comparison with Field Observations}
\begin{figure}
\centering
\includegraphics[width = \linewidth]{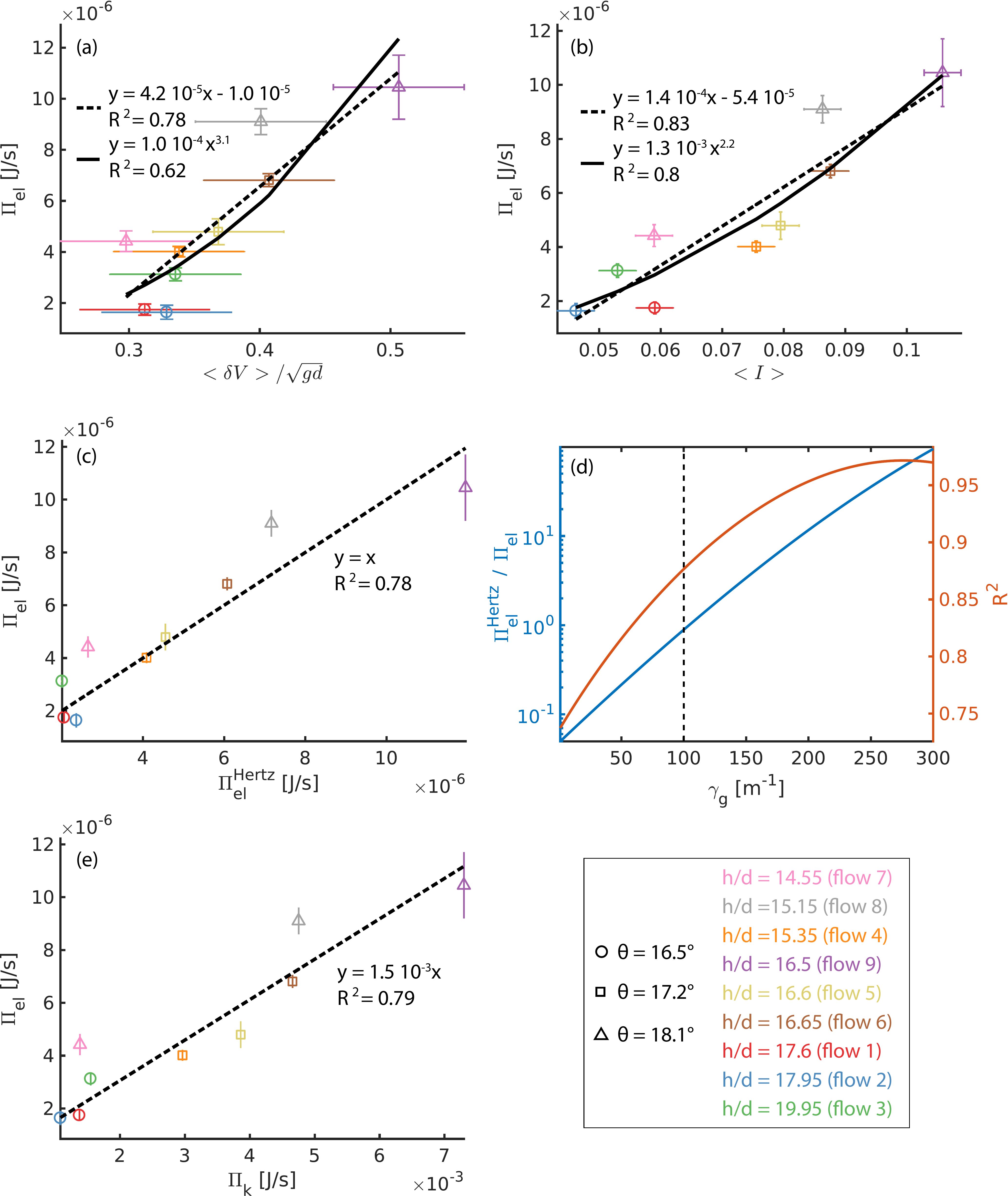}
	\caption{Radiated elastic power $\Pi_{el}$ as a function of (a) normalized average velocity fluctuations $<\delta V>/\sqrt{gd}$ and (b) average inertial number $<I>$. (c) Experimental $\Pi_{el}$ versus analytical elastic power $\Pi_{el}^{Hertz}$ for granular attenuation $\gamma_g = 100 \, \mathrm{m^{-1}}$.
Comparison with the line $y=x$ in red. (d) Model efficiency $\Pi_{el}^{Hertz}/\Pi_{el}$ calculated from the best linear fit between $\Pi_{el}^{Hertz}$ and $\Pi_{el}$ for a given attenuation as a function of the attenuation coefficient $\gamma_g$. The vertical black dashed line highlights
the case of $\gamma_g=100\,\mathrm{m^{-1}}$, the value for which the model gives about the same result as the measurements, i.e. $\Pi_{el}^{Hertz}/\Pi_{el}\simeq 1$.
(e) Comparison between the measured radiated elastic power $\Pi_{el}$ and analytical kinetic power $\Pi_k$.}
	\label{fig:Piel_scalings}
\end{figure}

We investigate here the relationship between the acoustic power and the properties of the flow averaged over the granular depth.
Figs. \ref{fig:Piel_scalings}(a) and (b) show that the acoustic power increases with the velocity fluctuations and the inertial number.
Our data are compatible with affine or power law relationships.
The best power laws that fit the data are
\begin{equation}
\Pi_{el} \propto \delta V^{3.1}  \propto I^{2.2}.
\label{scalingPiel}
\end{equation}
As velocity fluctuations are related to mean velocity by (Fig. \ref{fig:deltaV_gamma_I}d)
\begin{equation}
\delta V = 0.28 <V_x> + 0.08,
\end{equation}
the seismic power also scales as $\Pi_{el} \propto <V_x>^{3.1}$.

\begin{figure}
\centering
\includegraphics[width = \linewidth]{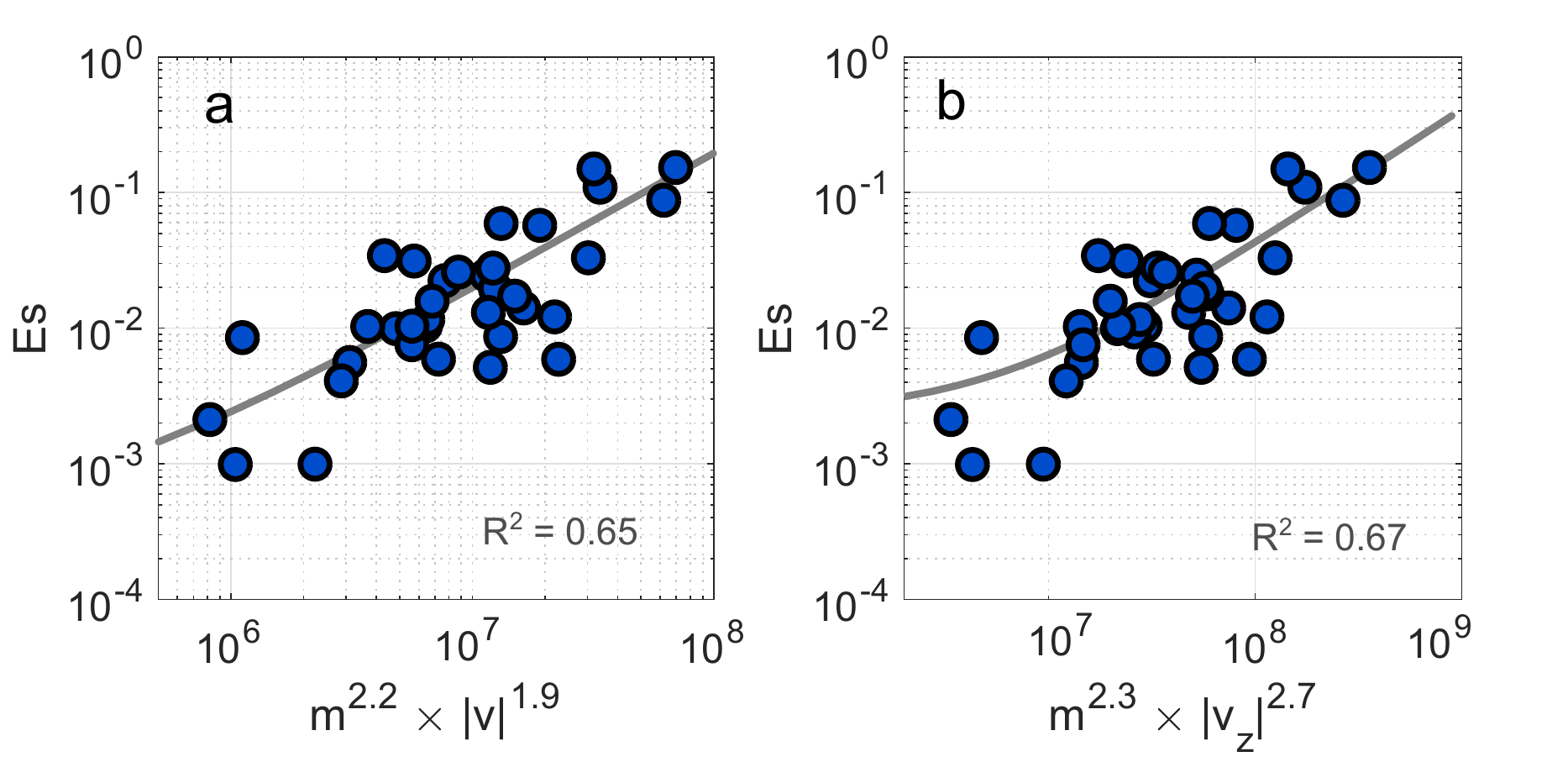}
	\caption{ a) Energy of the seismic signal generated at each individual block impact as a function of the block mass $m^\alpha$ and the modulus of the velocity before impact $v^\beta$, with the values of $\alpha$ and $\beta$ inferred to get the best fit by linear regression between those quantities; b) Similar derivation as a), considering only the vertical component of the velocity $v_z$.}
	\label{fig:Piel_fieldblocks}
\end{figure}

In the field, the seismic power calculated from the signal measured at seismic stations can be related to the mean flow velocity deduced from inversion of low-frequency seismic data \citep{Allstadt2013,Hibert2017b}.
Field experiments consisting in the release of single blocks of different masses have also shown a correlation between the velocity of the block before impact $v$ and the seismic energy $E_s$ released at the source \citep{Hibert2017c}. From this dataset, we looked for the exponents $\alpha$ and $\beta$ giving the best fitting regression line between $E_s$ and $m^\alpha \times v^\beta$, where $m$ is the mass of the block and $v$ the velocity before impact. When considering the modulus of the velocity, we found that the seismic energy scales as $E_s \propto |v|^{1.9}$ (Figure \ref{fig:Piel_fieldblocks}a). When considering only the modulus of vertical component of the velocity before impact $v_z$, the seismic energy scales as  $E_s \propto |v_z|^{2.9}$  (Figure \ref{fig:Piel_fieldblocks}b). These exponent values are not so far from those of our laboratory measurements, Eq.~(\ref{scalingPiel}), even though they were obtained for single blocks and not for granular flows. Note that similar scaling laws linking the seismic wave characteristics to the dynamic properties have been found for granular flows and single blocks for natural events (e.g. \citep{Schneider2010, Hibert2017b, Hibert2017c}). 

\subsubsection{Simple Model for Acoustic Emission}

We propose a simple model making it possible to recover the radiated elastic power from the velocity fluctuations (i.e. square root of the granular temperature) of the particles, based on the understanding of the seismic source gained above. We assume that (i) the elastic waves are generated during binary collisions between particles of adjacent layers at speeds corresponding to the particles fluctuation velocities, (ii) collisions are described by the Hertz contact law and the radiated elastic energy is the work done by the impact force during the contact \citep{Johnson1987, Farin2015}, and (iii) the acoustic waves propagate from the layer where they are generated down to the bottom of the channel, with attenuation $\gamma_g$.

The attenuation in granular material varies strongly depending on the confining pressure, packing fraction, etc. Different values are reported in the literature varying between $15-150\,\mathrm{m^{-1}}$: e.g. \citet{Voronina2004} 
found $\gamma_g = 100\,\mathrm{m^{-1}}$ and \citet{Hostler2005} found values between $25\,\mathrm{m^{-1}}$ and $50\,\mathrm{m^{-1}}$. The total analytical elastic power is obtained by summing up the contributions of all layers:
\begin{equation}
\label{eq:Piel_synth_raw}
\Pi_{el}^{Hertz} = \sum_{i=1}^n N_i \, W_{el, Hertz}^i \, e^{-\gamma_g y_i}
\end{equation}
where $W_{el, Hertz}^i$ is the elastic energy radiated during the impact of a particle of the layer $i$, $y_i$ the height of the center of the layer $i$, $e^{-\gamma_g y_i}$ the exponential decay of the wave energy with depth, $N_i$ the number of impacts per time unit in the layer $i$ and $n$ the number of layers.

The elastic energy radiated during an impact is computed from Hertz contact theory \citep{Farin2015} for impacts on plates
\begin{equation}
\label{eq:Wel_synth_raw}
W_{el, Hertz}^i = a_0 \left(\frac{d}{2}\right)^5 {\left(\delta V(y_i)\right)}^{11/5}
\end{equation}
with $\delta V(y_i)$, the velocity fluctuation in the layer $i$, and $a_0$, a prefactor involving the elastic parameters of the particles and the PMMA plate \citep{Bachelet2018}
\begin{equation}
\label{eq:Wel_synth_raw_a0}
a_0 \approx 2.1 \, \frac{1}{\sqrt{B \rho_p h_p}} \, \left( \frac{{E_b}}{2\,(1-\nu_b^2)} \,\rho_b^4 \right)^{2/5} \approx 1.4 \times 10^8 \,\mathrm{(I.S.U.)}
\end{equation}
The number of impacts per time unit in layer $i$ is given by:
\begin{equation}
\label{eq:Ni_strasbourg}
N_i = \frac{\phi l L}{\pi {(\frac{d}{2})}^2} f_i
\end{equation}
with the first term corresponding to the number of particles above the plate isolated acoustically and $f_i$, the number of impacts per particle and per time unit. Impacts are assumed to occur when a particle overrides another particle of the layer below at their relative downslope velocity
\begin{equation}
\label{eq:f_impacts}
f_i = \frac{V_x(y_i)-V_x(y_{i-1})}{d} = \dot{\gamma}(y_i)
\end{equation}
Combining expressions (\ref{eq:Piel_synth_raw}), (\ref{eq:Wel_synth_raw}), (\ref{eq:Ni_strasbourg}) and (\ref{eq:f_impacts}) leads to the final expression of the analytical radiated elastic power
\begin{equation}
\label{eq:Piel_synth2}
\Pi_{el}^{Hertz} = \frac{a_0 \phi l L}{8 \pi} \, d^3 \, \sum_{i} \dot{\gamma}(y_i) {\delta V(y_i)}^{11/5}  e^{-\gamma_g y_i}
\end{equation}
Using Eq. (\ref{eq:f_impacts}), the acoustic power is expected to scale as
\begin{equation}
\label{eq:Piel_scaling}
\Pi_{el} \propto {\delta V}^{16/5\simeq 3.2} \propto \dot{\gamma}^{3.2} \propto I^{1.6},
\end{equation}
as our optical observations showed that $\delta V \propto \left(\dot{\gamma}d\right) \propto I^{0.5}$. This is in very good agreement with the scaling
observed in Fig. \ref{fig:Piel_scalings}a that suggests $\Pi_{el} \propto {\delta V}^{3.1} \propto \dot{\gamma}^{3.1} \propto I^{1.55}$ even though, as said above,
the narrow range of our experiments makes it very difficult to discriminate between the affine and the power law relationship.

To compare our observations with those of \citet{Taylor2017}, we have to multiply what they called acoustic energy $E_a$ by $\omega^2$,
where $\omega$ is the pulsation because their definition of energy is a term proportional to the square of the acceleration, rather than the square of velocity.
As a result, to determine their energy, we have to multiply our scaling by $\delta V^{2/5}$ owing to Eq. (\ref{eq:fmean|hertz}).
This would lead to $E_a \propto {\delta V}^{18/5\simeq 3.6}$ and possibly $E_a \propto I^{2}$, if we assume the same relationship between
the mean quantities in their experiments. Their observations however suggest that $E_a \propto I$. This difference may be due to the fact that their setting is very
different from ours, to the uncertainty of their calculation of $I$ (i.e. they calculate $\dot{\gamma}$ by dividing the imposed velocity of the
shear cell by $5d$ for all experiments) or to the limitations of our simple model.

\begin{figure}
\centering
\includegraphics[width=12cm]{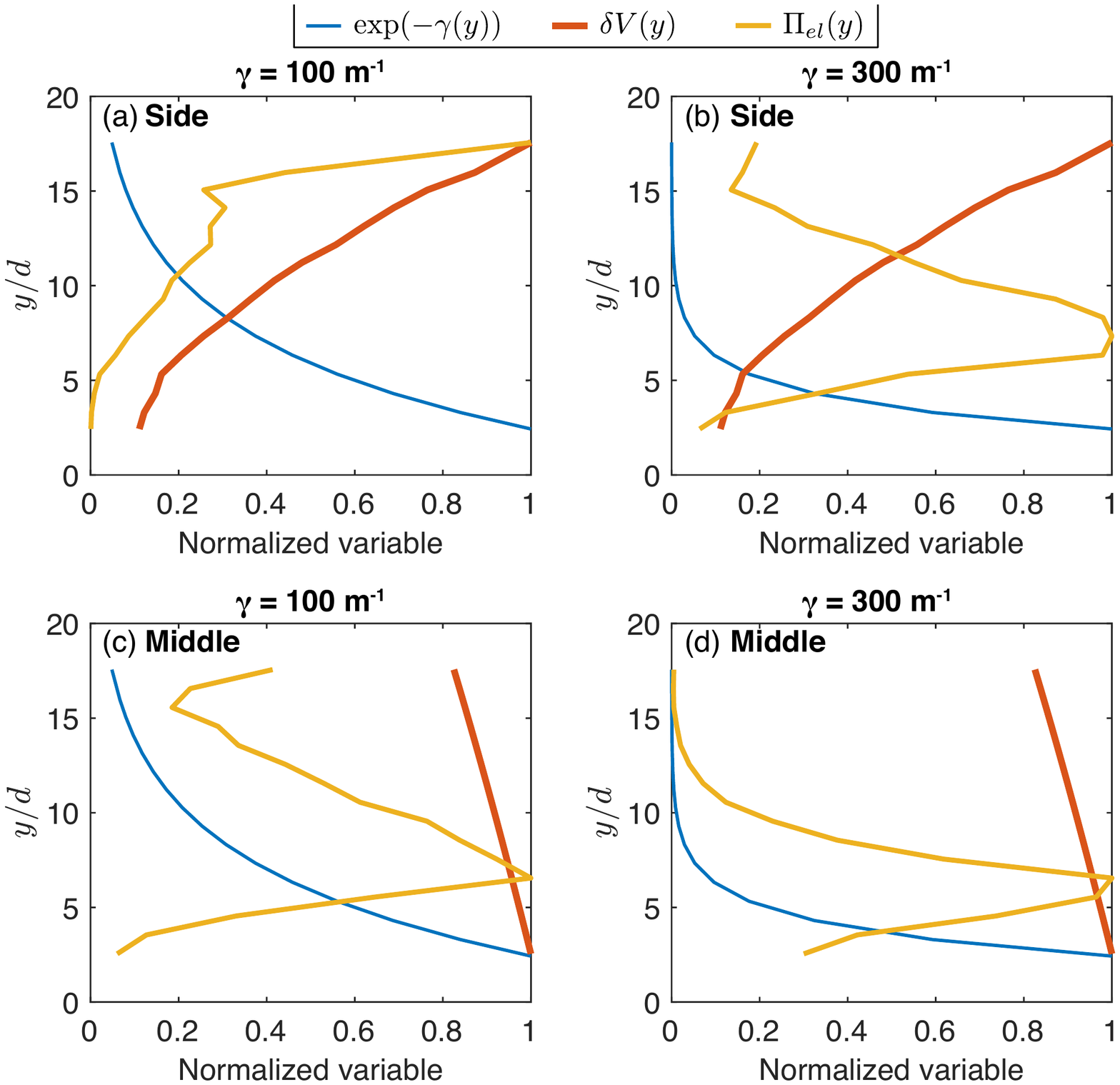}
	\caption{(a),(b) Analytical acoustic power $\Pi_{el}(y_i)$ per layer (i.e. as a function of depth $y/d$) computed using the fluctuating speed $\delta V$ measured along the side of the flow in experiment 1, for (a) $\gamma_g=100\,\mathrm{m^{-1}}$ and (b) $\gamma_g=300\,\mathrm{m^{-1}}$.
	(c),(d) Analytical acoustic power $\Pi_{el}(y_i)$ per layer computed assuming a linear granular temperature profile increasing with depth, as might be observed in the middle of the flow, for (c) $\gamma_g=100\,\mathrm{m^{-1}}$ and (d) $\gamma_g=300\,\mathrm{m^{-1}}$. In each panel, attenuation $\exp(-\gamma_g y)$ is also represented.}
	\label{fig:Piel_synth_layers}
\end{figure}

The key parameter in the calculation of $\Pi_{el}^{Hertz}$ is the attenuation factor. If we take $\gamma_g = 100\,\mathrm{m^{-1}}$, we obtain a very good
agreement with the measured acoustic power (Fig. \ref{fig:Piel_scalings}c). However, the value of $\Pi_{el}^{Hertz}$ is very sensitive to $\gamma_g$ as shown
in Fig. \ref{fig:Piel_scalings}d. For example if $\gamma_g = 50\,\mathrm{m^{-1}}$, $\Pi_{el}^{Hertz}\simeq 0.5 \Pi_{el}$. Figs. \ref{fig:Piel_synth_layers}(a) and (b)
show that with $\gamma_g = 100\,\mathrm{m^{-1}}$, the main contribution to the acoustic power comes from the grains near the surface, while
with $\gamma_g = 300\,\mathrm{m^{-1}}$, it comes from the grains located in the middle of the granular layer where velocities and velocity fluctuations are small.
Because the measured mean frequencies are closer to those associated with the highest velocity fluctuations within the flow (see section \ref{compare-frequency}), a larger contribution
of the grains near the surface seems more probable. This suggests a granular attenuation closer to $100\,\mathrm{m^{-1}}$.
Precise attenuation measurements will be a crucial step to further validate our simple model and will be performed in the future.

Another key issue is the difference between the fluctuations measured near the side walls and those within the flow as shown in the discrete element
simulations of \citet{Hanes2000} and discussed in section \ref{fluct-heterogen}. To assess the change in the acoustic power calculation if measurements
were performed in the flow center, we calculate $\Pi_{el}^{Hertz}$ by taking the same value $\delta V_{xs}$ of the fluctuating velocity $\delta V$ at the free surface and by assuming that $\delta V$
increases (instead of decreases) linearly down to the bottom to reach $\delta V(d) = 1.2\delta V_{xs}$ to mimic the simulations of \citet{Hanes2000} (their Fig. 15).
This assumption corresponds to
\begin{equation}
\label{eq:Piel_synth3}
\Pi_{el}^{Hertz}
= \sum_i \frac{\phi l L}{\pi (d/2)^2} f_i a_0 (d/2)^5 (1.2 \delta V_{xs} (1-y_i/h) + \delta V_{xs} y_i/h)^{11/5} e^{-\gamma_g y_i}
\end{equation}
Assuming that the collision frequency is $f_i=\delta V_i/d = (1.2 \delta V_{xs} (1-y_i/h) + \delta V_{xs} y_i/h) /d$  further leads to
\begin{equation}
\label{eq:Piel_synth4}
\Pi_{el}^{Hertz}
= \frac{a_0 \phi l L h d}{8 \pi} \delta V_{xs}^{16/5} \int_0^1 (1.2-0.2 z)^{16/5} e^{-\gamma_g h z} dz
\end{equation}
Note that when we make this assumption on the $\delta V(y)$ profile, the main contribution to the acoustic power comes from slightly below the middle of the granular layer, regardless of the attenuation coefficient $\gamma_g = 100\,\mathrm{m^{-1}}$ or $\gamma_g = 300\,\mathrm{m^{-1}}$ (Figs. \ref{fig:Piel_synth_layers}(c) and (d)).


\subsubsection{Acoustic versus Kinetic Energy}

Finally, in order to quantify the part of the kinetic energy of the flow converted into elastic energy, we assume that the conversion coefficient from kinetic to elastic energy, i.e.  the energy ratio or acoustic efficiency, $W_{el}/E_k=\xi$ generated by each impact is the same. We then replace the term $W_{el, Hertz}^i$ in expression (\ref{eq:Piel_synth_raw}) by $\xi \, E_k^i$, where $E_k^i$ is the average kinetic energy of particles of layer number $i$:
\begin{equation}
\label{eq:Piel_synth_kinetic_raw}
\Pi_{el}^{Hertz} = \xi \, \Pi_k = \xi \, \sum_{i} N_i \, E_k^i \, e^{-\gamma_g y_i}
\end{equation}
Fig. \ref{fig:Piel_scalings}e shows that the measured acoustic power varies as an affine function of $\Pi_k$ and that the energy ratio is $\xi=0.15\times 10^{-4}$.
For individual particle impacts, \citet{Bachelet2018} showed that the energy ratio $\xi$ decreases from about $10^{-1}$ for impacts on a rough bed to $10^{-3}$ for impacts on an erodible bed of a thickness equal to 10 particle diameters. As the highly energetic impacts are mostly at the surface here, at least near the side walls, we may consider that the particle impacts a bed of $15d$. If we use the empirical relationship proposed by \citet{Bachelet2018}, $\xi=0.13 e^{-0.44e^*}$, where $e^*$ is the number of layers of grains of the erodible bed, we find in our case, for $e^*=15$, that $\xi=1.7\times 10^{-5}$, in agreement with our results from granular flows.
This supports the simple model proposed here.
The energy ratio of $1.7\times 10^{-5}$ is very similar to what is observed in the field for rockfalls. As an example, values of $\xi\simeq 10^{-5}-10^{-3}$ were found for rockfalls on La R{\'e}union Island \citep{Hibert2011}, on Montserrat Island \citep{Levy2015} and in the French Alps \citep{Deparis2008}.

\section{Conclusion}


As seismic waves generated by landslides are continuously recorded by seismic networks, detailed analysis of these signals provides a new way to collect data on the dynamics and rheology of natural flows. This is however only possible if quantitative relationships between the flow properties and the acoustic signal characteristics are established. 

In the experiments reported here, we provide new quantitative insights into the origin of the acoustic signals generated by almost steady and uniform granular flows. By measuring precisely and synchronously the flow and generated waves with optical and acoustic sensors respectively, we have identified the essential physical sources of the waves. We have shown that the high-frequency signal (tens of kHz for mm-size glass particles) corresponding to the mean frequency of the signal is essentially related to particle collisions. This mean frequency is shown to be roughly proportional to the inertial number $I$. The measured acoustic power is well reproduced quantitatively with a simple model of particle impacts using Hertz theory and involving the relative particle velocity corresponding to flow velocity fluctuations. As our experiments show that velocity fluctuations roughly correlate with the mean flow velocity, our results suggest that mean flow velocity and velocity fluctuations could be determined from measurements of the high-frequency seismic signal. The conversion coefficient from kinetic to elastic energy, i.e. the energy ratio or acoustic efficiency, is around $10^{-5}$ in the experiments. This values is in rough agreement with field measurements where values of $10^{-6}-10^{-3}$ have been found for the ratio between seismic energy and potential energy lost [e.g. Hibert et al., 2011, Levy et al., 2005], given that kinetic energy is typically one order of magnitude smaller than potential energy lost (see Figures 6(a), (b) of \citet{Farin2018}). 

More precisely, our results suggest that the emitted seismic power is proportional to the granular temperature (square of velocity fluctuations). Beyond the interpretation of the generated acoustic signal in terms of granular flow properties, the measurement of velocity fluctuations and their link with mean properties may help improve our understanding of the behavior of natural flows near boundaries. Indeed, \citep{Artoni2015}  suggested that velocity fluctuations are a key ingredient to be included in models describing dense granular flows in the vicinity of an interface and appear in scaling laws reproducing the effective friction at lateral walls. More specifically, force fluctuations related to velocity fluctuations may trigger slip events even if the system is globally below the slip threshold \citep{Artoni2015}. Furthermore, velocity fluctuations, i.e. granular temperature, is a key parameter of the kinetic theory. Its measurement in dense granular flows will help constrain attempts to extend this theory to dense granular flows \citep{Berzi2014}.

Finally, a thousand times lower frequency (tens of Hz) is also identified in the acoustic signal and is shown to correspond to the displacement of particles over one another, related to the relative motion of the grain layers. This seems to result from the quasi monodisperity of the particles involved in these experiments. Further studies should investigate the role of particle size and shape on the generated acoustic signals and extend the range of bed slopes (i.e. velocities) so as to be able to better discriminate scaling laws between the flow and acoustic signal quantities. 


\clearpage

%
%

\begin{notation}

\notation{$a_0'$} Coefficient depending on elastic parameters (I.S.U.) (see Eq. \ref{eq_a0})
\notation{$a_z$} Vibration acceleration of the plate (m~s$^{-2}$)
\notation{$B$} Bending stiffness (J)
\notation{$\tilde{A}_z$} Amplitude spectrum of the vibration acceleration (m~s$^{-2}$/Hz)
\notation{$E_s$, $E_p$, $E^*$} Young's moduli (Pa)
\notation{$E_a$} Acoustic energy (m$^2$ s$^{-2}$) \citep[from][]{Taylor2017}
\notation{$E_k$} Kinetic energy (J)
\notation{$f$} Frequency of the vibration signal (Hz)
\notation{$f_i$} Number of impacts per particle and per time unit (s$^{-1}$)
\notation{$f_{mean}$} Mean frequency (Hz) (see Eq. (\ref{eq:fmean_exp}))
\notation{$f_{Hertz}$} Theoretical mean frequency predicted by the Hertz impact model (Hz) (see Eq. (\ref{eq:fmean|hertz}))
\notation{$f_{mod}$, $f_h$, $f_{flow}$, $f_v$} Characteristic frequencies generated by the granular flow (Hz) (see section \ref{section_discussion_freq})
\notation{$g$} Gravitational acceleration (m~s$^{-2}$)
\notation{$d$} Diameter of the particles (m)
\notation{$h$} Flow thickness (m)
\notation{$h_g$} Gate elevation (m)
\notation{$h_p$} Plate thickness (m)
\notation{$I$} Inertial number (-) (see Eq. (\ref{eq_Inertial_number}))
\notation{$L$, $l$} Dimensions of the acoustically isolated plate (m)
\notation{$M$} Mass of the acoustically isolated plate (g)
\notation{$N_i$} Number of impacts per unit time in particle layer $i$ (s$^{-1}$)
\notation{$n$} Number of particle layers (-)
\notation{$P$} Hydrostatic pressure (Pa)
\notation{$T$} Period of the signal (s)
\notation{$V_{x}$, $V_{y}$} Downslope and normal speeds of the particles (m~s$^{-1}$)
\notation{$<V_{x}>$, $<V_{y}>$} Speeds averaged within one layer (m~s$^{-1}$)
\notation{$V_{xs}$} Downslope speed of the surface particles (m~s$^{-1}$)
\notation{$v_{g}$} Group speed of $A_0$ mode in PMMA $v_g \approx 1000$ (m~s$^{-1}$)
\notation{$v_{z}$} Vibration speed of the plate (m~s$^{-1}$)
\notation{$x$, $y$} Downslope and normal positions of the particles (m)
\notation{$w$} Thickness of one layer of particles in the flow (m)
\notation{$W_{el}$} Radiated elastic energy (J)
\notation{$\dot{\gamma}$} Shear rate (s$^{-1}$)
\notation{$\gamma_g$} Characteristic attenuation coefficient of acoustic energy in granular media (m$^{-1}$)
\notation{$\Delta t$} Duration (s)
\notation{$\delta V_x$, $\delta V_y$} Downslope and normal fluctuating speeds of particles (m~s$^{-1}$)
\notation{$\theta$} Slope angle ($^\circ$)
\notation{$\nu_p$, $\nu_s$} Poisson's ratios (-)
\notation{$\xi$} Energy ratio, acoustic efficiency (-)
\notation{$\Pi_{el}$} Radiated elastic power (J s$^{-1}$)
\notation{$\Pi_{el}^{Hertz}$} analytical radiated elastic power (J s$^{-1}$)
\notation{$\Pi_{k}$} Kinetic power (J s$^{-1}$)
\notation{$\rho$, $\rho_p$, $\rho_s$} Densities (kg m$^3$)
\notation{$\phi$} Packing fraction ($-$)
\notation{$\omega$} Pulsation (s$^{-1}$)

\end{notation}

\section{Acknowledgments}
We thank Xiaoping Jia, Sylvain Viroulet, Diego Berzi and Alexandre Valance for insightful discussions on  granular temperature and kinetic theory. We thank  G{\"o}ran Ekstr{\"o}m for sharing the inverted force history for the Mt Dall, Mt Lituya, Sheemahant Glacier and Lamplughr Glacier landslides. We thank Alain Steyer for his great help in mounting the setup. RT acknowledges the support of the INSU ALEAS and the France-Norway LIA D-FFRACT programs.
The data acquired during the experiments and the scripts to treat them are available on the repository \citep{Bachelet2020}.


\appendix



\section{Heights of the Flows}

\begin{figure}
\centering
\includegraphics[width = \linewidth]{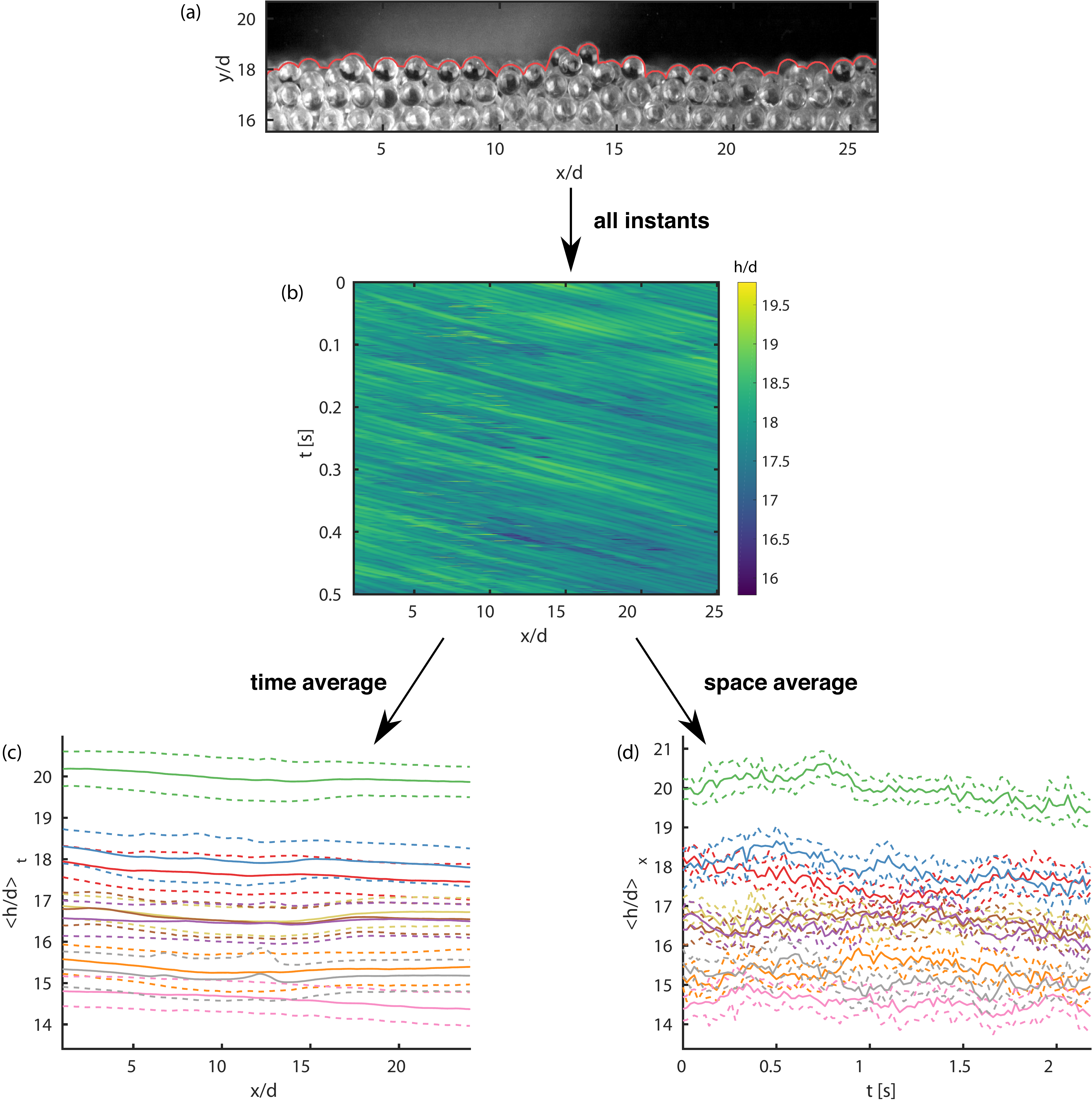}
	\caption{Heights of the flows: (a) example of flow interface detection (red line), (b) space and time height, thereafter averaged over (c) time or (d) space. Each color of panels (c) and (d) corresponds to a specific flow (see for example Fig. \ref{fig:Piel_scalings} for detailed legend).}
	\label{fig:Height_flows}
\end{figure}

The flow height is measured by tracking the particles at the free surface of the flow (Fig. \ref{fig:Height_flows}a) (procedure similar to that used for particle tracking). Then, the spatial and temporal height obtained by repeating the procedure for all instants (Fig. \ref{fig:Height_flows}b) is averaged over time (Fig. \ref{fig:Height_flows}c) and space (Fig. \ref{fig:Height_flows}d).

\section{Velocity Fluctuation Measurements: Window Effect}
\label{windowsize}

The estimate of total velocity fluctuations depends on the width $w$ of the window considered:
\begin{equation}
\label{eq:deltaV_window_raw}
\delta V^2(y,t) = \frac{1}{w} \, \int_{y-w/2}^{y+w/2} {\left( \mathbf{V}(y',t)-\left<\mathbf{V}\right>(y,t) \right)}^2 dy'
\end{equation}
where $\left<\mathbf{V}\right>(y,t)$ is the average velocity in the center of the box. A first order expansion $\left<\mathbf{V}\right>(y,t) = \left<\mathbf{V}\right>(y',t) - \dot{\gamma}(y) (y'-y) \mathbf{e_x}$ gives (the average vertical velocity equals zero):
\begin{equation}
\delta V^2(y,t) = \frac{1}{w} \, \int_{y-w/2}^{y+w/2} {\left( \mathbf{\delta V}^*(y') + \dot{\gamma}(y) (y'-y) \mathbf{e_x} \right)}^2 dy'
\end{equation}
with $\mathbf{\delta V}^*(y') = \mathbf{V}(y',t)-\left<\mathbf{V}\right>(y',t)$. Developing the square leads to three terms $I_1$, $I_2$ and $I_3$:
\begin{equation}
I_1 = {\delta V^*}^2(y,t) = \frac{1}{w} \, \int_{y-w/2}^{y+w/2} {\mathbf{\delta V}^*}^2(y') dy'
\end{equation}
\begin{equation}
I_2 = \frac{2}{w} \, \int_{y-w/2}^{y+w/2} \dot{\gamma}(y) (y'-y) \delta V_x(y') dy'
\end{equation}
\begin{equation}
I_3 = \frac{1}{w} \, \int_{y-w/2}^{y+w/2} {\left(\dot{\gamma}(y) (y'-y) \right)}^2 dy' = \frac{w^2 \, \dot{\gamma}^2(y)}{12}
\end{equation}
$I_1$ corresponds to the genuine velocity fluctuations averaged on the box. $I_2$ can be computed by a first order expansion of $\delta V_x(y')$:
\begin{equation}
\delta V_x(y') = \delta V_x(y) + \frac{d \delta V_x}{dy}(y) (y'-y)
\end{equation}
Thus:
\begin{equation}
I_2 = \frac{2}{w} \, \left( \delta V_x(y) \, \int_{y-w/2}^{y+w/2} (y'-y) dy' + \frac{d \delta V_x}{dy}(y) \, \int_{y-w/2}^{y+w/2} (y'-y)^2 dy' \right)
\end{equation}
The first term equals zero, whereas the second can be neglected because of the second order.

Finally, total velocity fluctuations estimate are given by the following expression:
\begin{equation}
\label{eq:deltaV_window_end}
\delta V^2(y,t) = {\delta V^*}^2(y,t) + \frac{w^2 \, \dot{\gamma}^2(y)}{12}
\end{equation}
The second term quantifies the error introduced by considering the average velocity taken in $y$ (the center of the box) instead of the value in $y'$ in formula (\ref{eq:deltaV_window_raw}). Its expression is very similar to the one found by \citet{Weinhart2013} (Eq. (34)). The only difference comes from the choice of the averaging function, also called the coarse-graining function. We implicitly chose a gate equal to one in $[y-w/2, y+w/2]$ and to zero elsewhere, whereas a more complex choice is usually selected for differentiability \citep{Glasser2001, Weinhart2013}.

\begin{figure}
\centering
\includegraphics[width = 8cm]{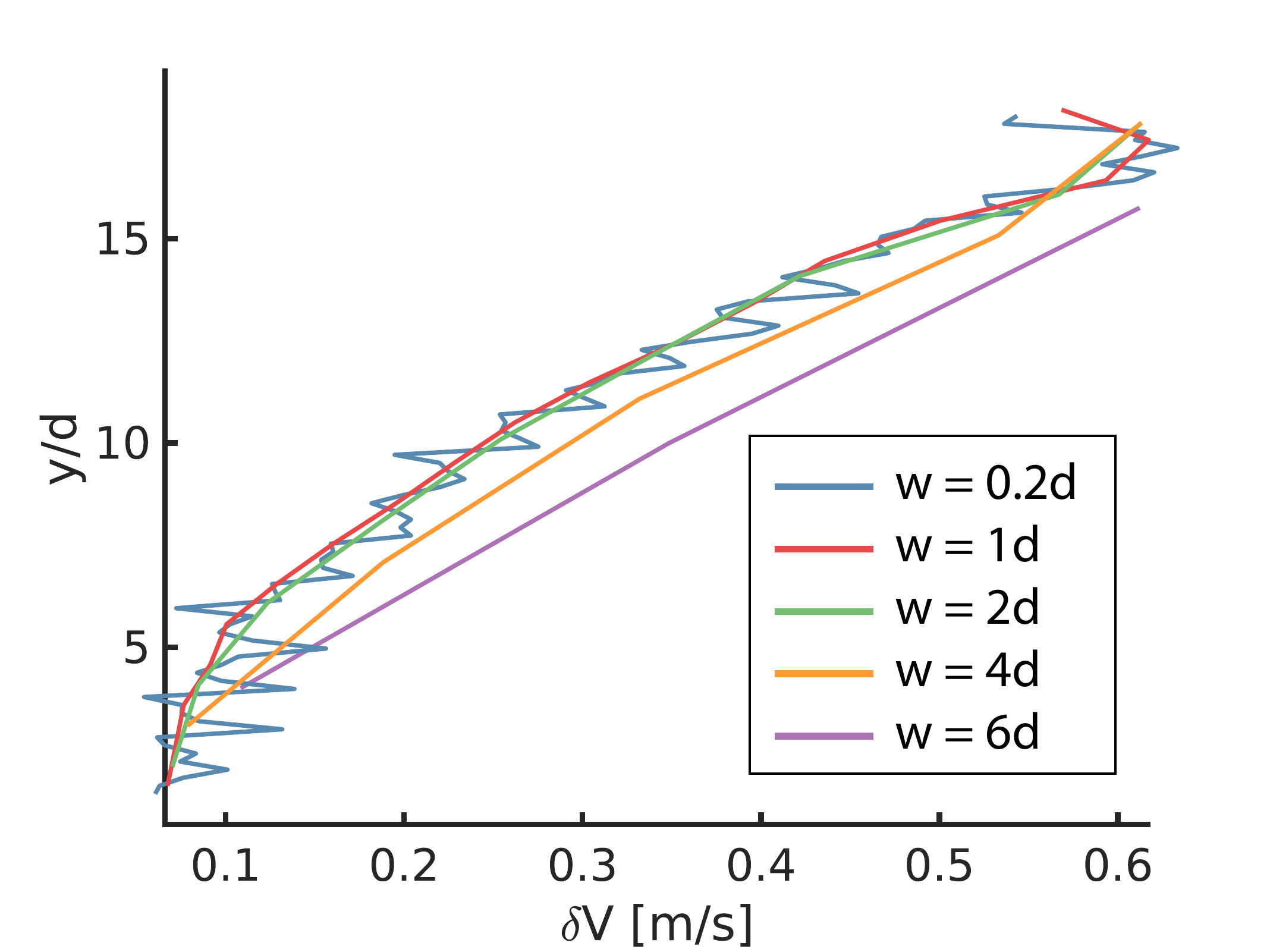}
	\caption{Effect of the window size on the fluctuation velocity computation.}
	\label{fig:deltaV_window}
\end{figure}

Thanks to expression (\ref{eq:deltaV_window_end}) and approximating $\delta V^*$ by $2.1 \, d \, \dot{\gamma}$, as suggested by the linear fit in Fig. \ref{fig:deltaV_gamma_I}e, it is possible to deduce that the windows have an effect similar to that of $\delta V^*$ when $w=5d$. For this reason, the window is negligible in our case (see Fig. \ref{fig:deltaV_window})

\section{Correlation Lengths within the Flow}
\label{correlation}
To obtain quantitative measurements of the correlation length of velocity fluctuations we compute the
downslope and vertical velocity correlations between two points $M_1$ and $M_2$ with coordinates $(x_1, \, y_1)$ and $(x_2, \, y_2)$:
\begin{equation}
\label{eq:corr_u_t}
C_{V_i}(M_1,M_2) = \frac{\sum_{t} \delta V_i(M_1,t) \times \delta V_i(M_2,t)}{\sqrt{\sum_t \delta V_i(M_1,t)^2} \times \sqrt{\sum_t \delta V_i(M_2,t)^2}}
\end{equation}
where $i=x, y$. Examples of downslope and vertical velocity correlations are presented in Figs. \ref{fig:Correlations_all}(a) and (b) respectively. High correlations of the horizontal velocity over one particle thickness are clearly visible. To quantify this correlation, a correlation length has been defined. It corresponds to the length at which the correlation reaches a given threshold. Unlike  \citet{Pouliquen2004} who chose a threshold of $0.05$, we selected a value of $0.5$ because of the limitation of the window of observation (see the dark grey contour plot of Fig. \ref{fig:Correlations_all}a which seems cropped by the right border of the window). The correlation length increases with decreasing slope angle as observed by \citet{Pouliquen2004} and \citet{Staron2008} or in granular flows approaching jamming \citep{Gardel2009}. In our experiments, only the lengths of downslope velocities in the $x$-direction $\lambda_{xx}$ are higher than one particle diameter. This suggests correlated motion of particles of the same layer, supporting the layering observed in Fig. \ref{fig:particle_tracking}b. In agreement with \citet{Pouliquen2004} and \citet{Staron2008}, correlation lengths decrease for increasing slope angles (Fig. \ref{fig:Correlations_all}c-e), as observed in Movies 3 and 4 (supplementary material). The correlation lengths collapse to zero under $y/d=5$ because particle velocities are smaller than noise.

Note that for dry granular chute flows \citep{Gardel2009} and for granular flows in a fluid \citep{Orpe2007}, significantly greater spatial correlations are observed near the boundaries, which may be the case here.

\begin{figure}
\centering
\includegraphics[width = 10cm]{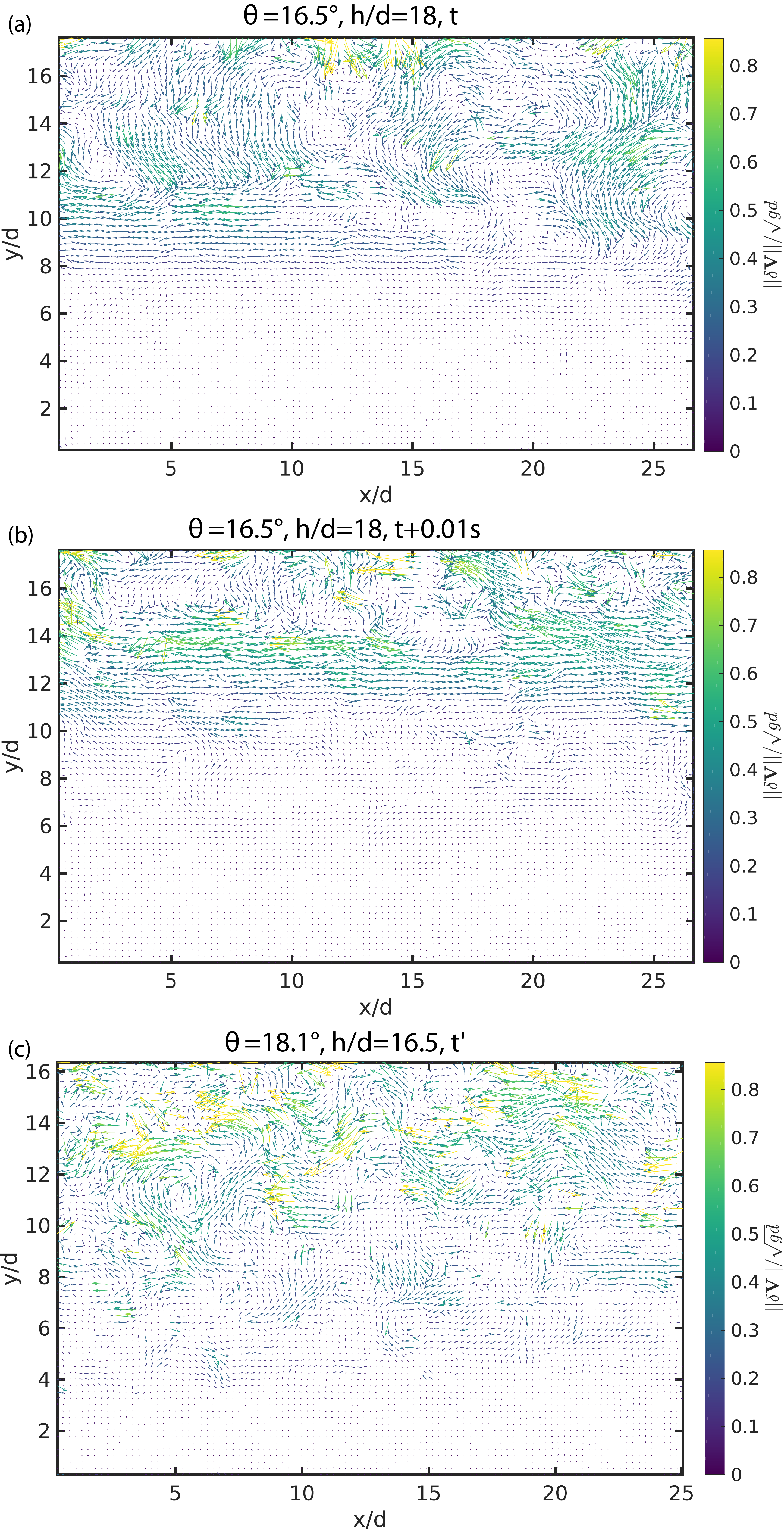}
	\caption{Map of velocity fluctuations obtained with CIV for flow number $2$ (a,b) and $9$ (c) at instants $t$, $t+0.01 \, \mathrm{s}$ and $t'$ respectively ($t$ and $t'$ are arbitrary).}
	\label{fig:map_deltaV}
\end{figure}

\begin{figure}
\centering
\includegraphics[width = 12cm]{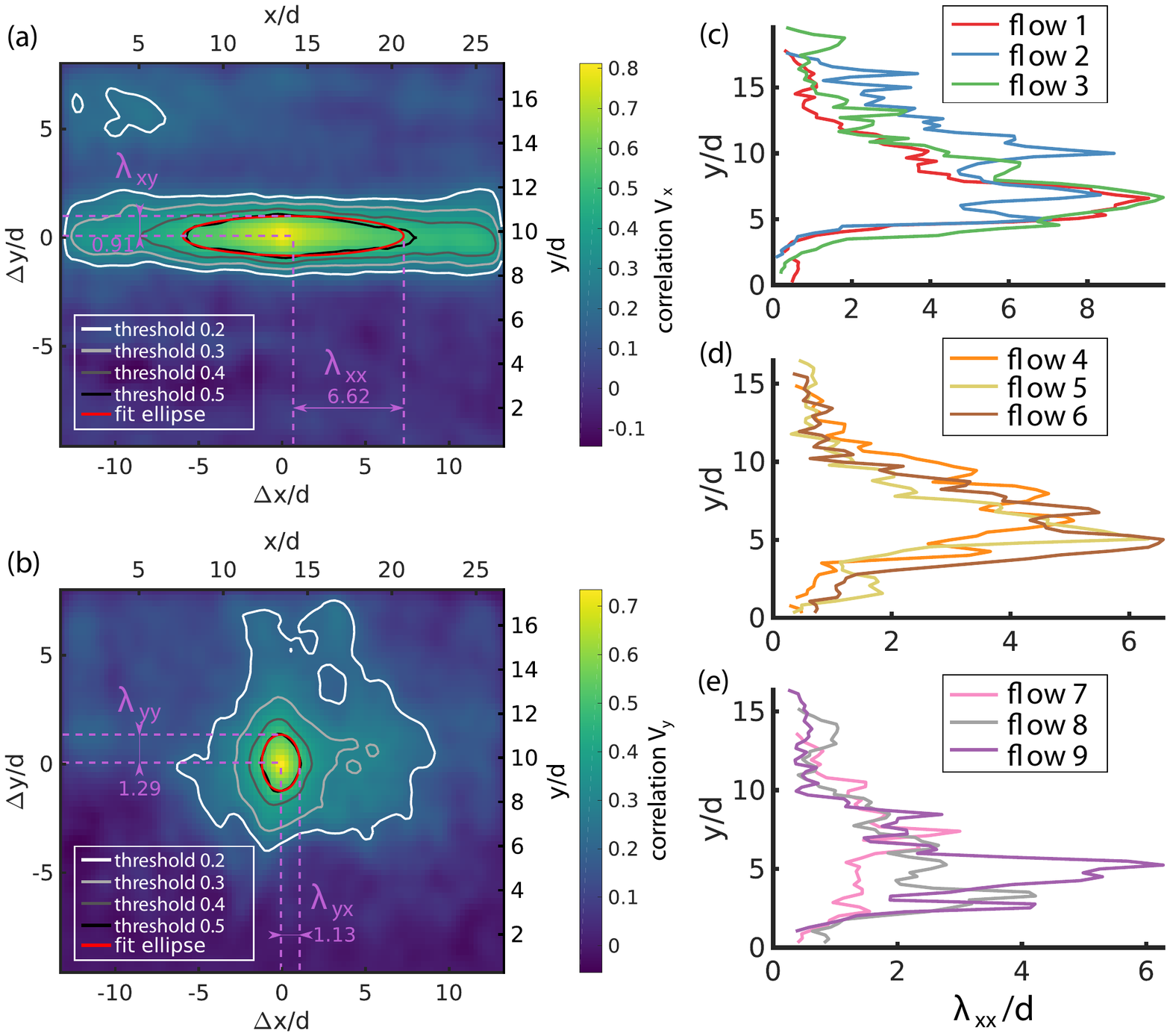}
	\caption{Example of (a) horizontal and (b) vertical velocity correlations between the static point of coordinates ($x/d=14$, $y/d=10$) and all the others positions for the flow $2$. Panels (c) to (e) correspond to the correlation length $\lambda_{xx}$ of the horizontal velocity in x direction for all the flows.}
	\label{fig:Correlations_all}
\end{figure}

\begin{figure}
\centering
\includegraphics[width = 8cm]{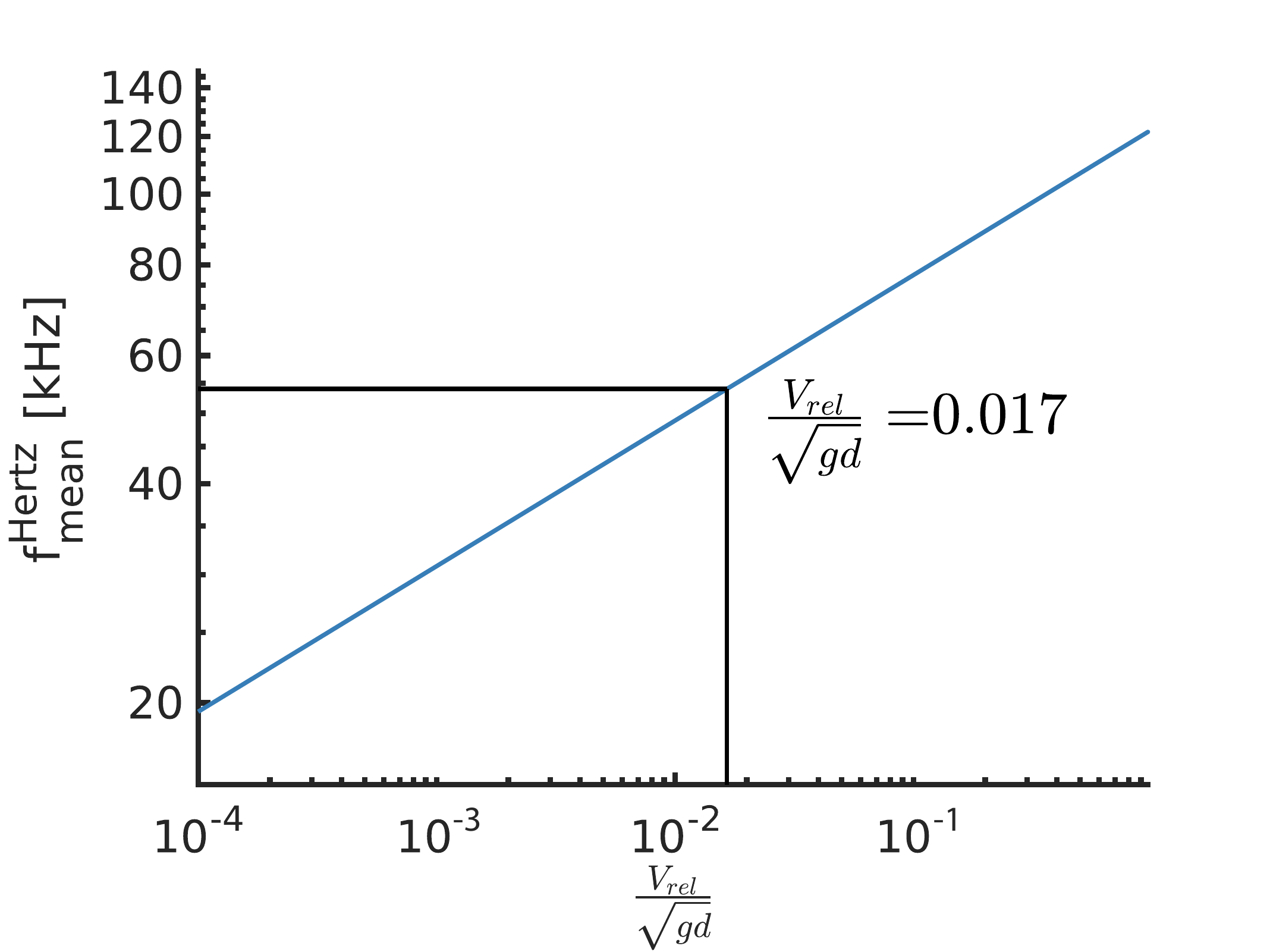}
	\caption{Average frequencies generated by impacts between two glass particles at relative velocity $V_{rel}$. The maximum velocity associated with the maximal frequency of $54 \, \mathrm{kHz}$ of our sensors is also presented.}
	\label{fig:fmean_Hertz}
\end{figure}

\end{document}